\documentclass[12pt]{article}

\usepackage{epsfig}
\include{amssym}

\def\mathbf{\vec}

\def\ca{\c{c}\~{a}}

\newcommand{\eff}{\mathrm{ef\mbox{}f}}
\newcommand{\ud}{\mathrm{d}}
\newcommand{\p}{\mathrm{p}}
\newcommand{\id}{\mathrm{id}}
\newcommand{\s}{\mathrm{s}}

\begin{document}

\centerline{\Large\bf\sc Effects of eight-quark interactions} 
\vspace{0.2cm}
\centerline{\Large\bf\sc on the hadronic vacuum and mass spectra} 
\vspace{0.2cm}
\centerline{\Large\bf\sc of light mesons} 
\vspace{1cm}

\centerline{\large A. A. Osipov\footnote{On leave from Joint 
            Institute for Nuclear Research, Laboratory of Nuclear 
            Problems, 141980 Dubna, Moscow Region, Russia. 
            Email address: osipov@nusun.jinr.ru}, 
            B. Hiller\footnote{Email address: 
            brigitte@teor.fis.uc.pt},
            A. H. Blin\footnote{Email address: alex@teor.fis.uc.pt},
            J. da Provid\^encia\footnote{Email address:
            providencia@teor.fis.uc.pt}} 
\vspace{0.5cm}
\centerline{\small\it Centro de F\'{\i}sica Te\'{o}rica, 
            Departamento de
            F\'{\i}sica da Universidade de Coimbra,}
\centerline{\small\it 3004-516 Coimbra, Portugal}
\vspace{1.0cm}

\centerline{\bf Abstract}
\vspace{0.5cm}
The combined ef\mbox{}fective low energy QCD Lagrangians of Nambu -- 
Jona-Lasinio (NJL) and 't Hooft are supplemented with eight-quark 
interactions. This work is a follow-up of recent f\mbox{}indings, 
namely (i) the six quark f\mbox{}lavour determinant 't Hooft term 
destabilizes the NJL vacuum, (ii) the addition of a chiral invariant 
eight-fermion contact term renders the ground state of the theory 
globally stable; (iii) stability constrains the values of coupling 
constants of the model, meaning that even in the presence of 
eight-quark forces the system can be unstable in a certain parameter
region. In the present work we study a phenomenological output 
of eight-quark interactions considering the mass spectra of 
pseudoscalar and scalar mesons. Mixing angles are obtained and their 
equivalence to the two angle approach is derived. We show that the
masses of pseudoscalars are almost neutral to the eight-quark forces. 
The only marked effect of the second order in the $SU(3)$ breaking is 
found in the $\eta -\eta'$ system. The scalars are more sensitive to 
the eight-quark interactions. A strong repulsion between the 
singlet-octet members is the reason for the obtained low mass of the 
$\sigma$ state within the model considered.

\vspace{1.0cm}
\noindent
PACS number(s): 12.39.Fe, 11.30.Rd, 11.30.Qc

\newpage 
%%%%%%%%%%%%%%%%%%%%%%%%%%%%%%%%%%%%%%%%%%%%%%%%%%%%%%%%%%%%%%%%%%%%%%%%%%%%%%

\section*{\centerline{\large\bf 1. Introduction}}

%%%%%%%%%%%%%%%%%%%%%%%%%%%%%%%%%%%%%%%%%%%%%%%%%%%%%%%%%%%%%%%%%%%%%%%%%%%%%%
The fundamental fields of QCD, quarks and gluons, are unobservable
dynamical variables. Instead, at low energies, one observes hadrons. 
The most direct way to study their properties is the method of
effective Lagrangians written in terms of the matter fields describing
mesons or baryons. Such theories, under some circumstances, can be 
developed to the advanced level of an effective field theory. A 
well-known example is chiral perturbation theory \cite{Gasser:1984}, 
where the Lagrangian of light pseudoscalar mesons is both a 
derivative and light current quark mass expansion around the 
asymmetric ground state which is assumed to be stable. This stability 
is a phenomenological fact; the underlying theory must explain it.

It is commonly accepted that the eight approximate Goldstone bosons
$\pi , K$ and $\eta$ are a signal for the spontaneous breakdown of 
the chiral symmetry of QCD, which is realized in the ideal world of 
massless $u, d, s,$ quarks. It is not excluded that effective
four-quark interactions of the Nambu and Jona-Lasinio (NJL) type 
\cite{Nambu:1961} are responsible for the formation of a stable chiral 
asymmetric vacuum giving a crude insight into the structure of the 
ground state of QCD \cite{Shuryak:2004}. 

One might ask if higher order quark interactions are of importance.  
For instance, on lines suggested by an instanton-gas model, it can be
argued \cite{Simonov:1997} that there exists an infinite set of 
multi-quark terms in the effective quark Lagrangian starting from the 
NJL four-quark interactions. The famous 't Hooft determinantal  
interaction \cite{Hooft:1978} automatically appears if one keeps only 
the zero mode contribution in the mode expansion of the effective 
Lagrangian. This $2N_f$ multi-quark term ($N_f$ being the number of quark 
flavours) manifestly violates the $U_A(1)$ axial symmetry of the QCD 
Lagrangian, offering a way out of the $U_A(1)$ problem.
  
Let us recall the case with the lightest flavour singlet pseudoscalar 
$\eta'$, which was for a while a deep theoretical problem, known as 
the $U_A(1)$ puzzle. The general solution \cite{Witten:1979} has 
shown that the $\eta'$, being a quark-antiquark state, is strongly 
connected to the gluon world and that the $U_A(1)$ axial anomaly is
the reason for the $\eta' -\pi ,K,\eta$ splitting observed in nature. 
These conclusions are based only on the Ward identities and the $1/N_c$
expansion of QCD (where $N_c$ is the number of colours). 

The same question has been also studied in the framework of an
effective Lagrangian which includes the meson fields and the
topological charge density $Q(x)$ \cite{Vecchia:1979}. After the 
elimination of the field $Q(x)$ by means of its classical equation 
of motion, one obtains an effective mesonic Lagrangian. It has been 
shown by Rosenzweig, Schechter and Trahern \cite{Schechter:1980} that 
the 't Hooft type determinantal interaction, written in terms of mesonic 
fields, appears as the first term in the expansion which results from  
eliminating $Q(x)$. From the Lagrangian of the model one realizes again 
that there are no valid theoretical objections against the idea that
the 't Hooft interaction and higher order multi-quark terms are
actually present in the QCD vacuum. Part of these~interactions have
been utilized in \cite{Alkofer:1989}.
   
Thus, it is tempting to consider the intuitive picture that describes
the QCD vacuum with basis on a series of multi-quark interactions  
reflecting several tractable features of QCD, which include aspects of 
chiral symmetry and of the $1/N_c$ expansion. The bosonization of quark 
degrees of freedom leads then to the desirable effective Lagrangian with
matter fields and a stable chiral asymmetric vacuum.  

This idea is not new. The NJL type model with the $U_A(1)$ axial
symmetry breaking by the 't Hooft determinant (in the following 
we will use the abbreviation NJLH for this model) has been studied
in the mean field approximation \cite{Bernard:1988}-\cite{Bernard:1993} 
for a long time. Numerous phenomenological applications show that the 
results of such an approach meet expectations. Nevertheless in this 
picture, there is an apparent problem: the mean field potential is 
unbounded from below, and the 't Hooft term is the direct source of
such an instability (see, for instance, Eq. (3.16) in \cite{Hatsuda:1994}). 
A consistent approach requires obviously a stable hadronic vacuum in
which the pions would live forever in the ideal world with only strong 
interactions.

The functional integral bosonization of the model exposes new 
shortcomings: the system of stationary phase equations used to
estimate the generating functional of the theory $Z$, has several real 
solutions \cite{Osipov:2005a} which contribute independently, i.e., 
$Z=Z_1+Z_2+\ldots +Z_n$, where $n$ is the total number of such real 
solutions. Since only one of them (let us assume $Z_1$, for 
definiteness) leads, at leading order, to the mean f\mbox{}ield 
potential, $V_{MF}$, the semiclassical potential $V$, corresponding to 
$Z$, dif\mbox{}fers from $V_{MF}$. It has been shown in 
\cite{Osipov:2005a} that $V$ is also unbounded from below. Thus, we
must accept that the NJLH model suffers from a ground state problem. 

Recently it has been argued \cite{Osipov:2005b} that eight-quark 
interactions, added to the NJLH Lagrangian, might resolve the problem. 
Indeed, the mean field potential of the modified theory, 
${\cal V}_{MF}$, has a globally stable minimum. The just mentioned 
controversy concerning the results obtained by the mean field method
and the functional integral approach is also removed: one can prove that 
${\cal V}={\cal V}_{MF}$, i.e., the number of admissible real solutions 
to the stationary phase equations can be constrained to one due to 
eight-quark terms. 

There is a natural question. If the eight-quark forces are so
important for the formation of the ground state, what are the other 
phenomenological consequences of such interactions?  

In this paper we consider the main characteristics of light
pseudoscalar mesons ($J^{PC}=0^{-+}$): their masses and weak decay 
constants. After that we switch to scalars, calculating masses of the 
$J^{PC}=0^{++}$ quark-antiquark nonet. The structure of scalars 
is a subject of many studies nowadays. The question is so complicated
that it would be too naive to think that eight-quark forces are a 
panacea for the mass spectrum problem. Our aim is only to demonstrate 
the tendency. Once we understand what is changed by the new
interactions considered in the description of the meson properties
within the model, we can clarify the role of eight-quark forces for 
low-energy QCD.  

To study this matter one should choose an appropriate approximation. 
The bosonization of six- and eight-quark interactions cannot be done 
exactly. We will use the stationary phase method to replace the 
multi-quark vertices by purely mesonic ones and by Yukawa type 
interactions of quarks with mesons. This is a standard approach 
\cite{Reinhardt:1988}, \cite{Eguchi:1976}-\cite{Diakonov:1996}. 
The subsequent integration over quarks is a straightforward
calculation, because one deals here with a Gaussian integral. To
obtain the effective mesonic Lagrangian and extract masses, we shall 
expand the real part of the quark determinant in a heat kernel series 
\cite{Schwinger:1951,Ball:1989}. The techniques which are particularly 
well suited to the present task have been developed in 
\cite{Osipov:2001}.  
 
The outline of the paper is as follows: after introducing the
multi-quark Lagrangian in Section 2.1, the stability conditions of the 
vacuum are discussed in Section 2.2. In Sections 2.3 - 2.5 we use 
bosonization and heat kernel methods to transform the multi-quark into 
a mesonic Lagrangian, and extract the relevant contributions to the
mass spectra and gap-equations. Section 3 is dedicated to the analysis
of the pseudoscalar observables and includes a detailed discussion of 
decay constants with particular emphasis on the relation of our one 
angle approach to the two mixing angle analysis. Explicit formulae for 
masses and mixing angles are obtained in Section 3.4. In Section 4 are 
presented the characteristics of scalars, numerical results are given
in Section 5 and conclusions in Section 6. Two appendices contain 
respectively the detailed derivation of the uniqueness of the
solutions of the stationary phase equations and of the solution of the 
equations which yield the matricial coefficients relevant for meson
mass terms.
%%%%%%%%%%%%%%%%%%%%%%%%%%%%%%%%%%%%%%%%%%%%%%%%%%%%%%%%%%%%%%%%%%%%%%%%%%%%%%

\section*{\centerline{\large\bf 2. The model}}

%%%%%%%%%%%%%%%%%%%%%%%%%%%%%%%%%%%%%%%%%%%%%%%%%%%%%%%%%%%%%%%%%%%%%%%%%%%%%%

\subsection*{\it 2.1 The multi-quark Lagrangian}

We discuss the system of light quarks $u, d, s$ $(N_f=3)$ with 
multi-fermion interactions described by the Lagrangian
\begin{equation}
\label{efflag}
  {\cal L}_\eff =\bar{q}(i\gamma^\mu\partial_\mu - m)q
          +{\cal L}_{4q} + {\cal L}_{6q}
          +{\cal L}_{8q}+\ldots\, .
\end{equation}
Quark f\mbox{}ields $q$ have colour $(N_c=3)$ and f\mbox{}lavour 
indices which are suppressed. We suppose that four-, six-, 
and eight-quark interactions ${\cal L}_{4q}$, ${\cal L}_{6q}$, 
${\cal L}_{8q}$ are ef\mbox{}fectively local. Likewise, they are
constructed from local quark bilinears, like the scalar $S_a=\bar q
\lambda_a q$, or the pseudoscalar $P_a=\bar q i\gamma_5\lambda_a q$ 
``currents''. Such bilinears have the appropriate quantum numbers to 
describe mesons. This approximation corresponds to the task
considered: we want to obtain, after bosonization, the tree level 
ef\mbox{}fective meson Lagrangian, with local vertices and local meson 
fields, and relate the coupling constants and masses of such a 
Lagrangian with the parameters of the quark model. Meson physics in
the large $N_c$ limit is described by a local Lagrangian of this type 
\cite{Witten:1979b}.

The global chiral $SU(3)_L\times SU(3)_R$ symmetry of the Lagrangian 
(\ref{efflag}) at $m=0$ is spontaneously broken to the $SU(3)$ group, 
showing the dynamical instability of the fully symmetric solutions of 
the theory. In addition, the current quark mass $m$, being a diagonal
matrix in f\mbox{}lavour space with elements $\mbox{diag} (m_u, m_d, 
m_s)$, explicitly breaks this symmetry down, retaining only the
reduced $SU(2)_I\times U(1)_Y$ symmetries of isospin and hypercharge 
conservation, if $m_u = m_d \neq m_s$. 

The leading order (in $N_c$ counting) Lagrangian of light mesons
and the corresponding underlying quark Lagrangian must inherit the
$U(3)_L\times U(3)_R$ chiral symmetry of massless three-flavour
QCD. In particular, it was argued \cite{Colwit:1980} that in the large 
$N_c$ limit of QCD with three massless quarks the pattern of spontaneous 
chiral symmetry breakdown is uniquely fixed: the chiral $U(3)_L\times 
U(3)_R$ group, under some highly plausible assumptions, necessarily 
breaks down to the diagonal $U(3)$. In accordance with these
expectations the short-range attractive $U(3)_L\times U(3)_R$
symmetric NJL-type interaction 
\begin{equation}
\label{L4q}
  {\cal L}_{4q} =\frac{G}{2}\left[(\bar{q}\lambda_aq)^2+
                 (\bar{q}i\gamma_5\lambda_aq)^2\right]
\end{equation}
can be used to specify the corresponding part of the ef\mbox{}fective 
quark Lagrangian in channels with quantum numbers $J^P=0^+, 0^-$ 
\cite{Volkov:1984}. The matrices acting in f\mbox{}lavour space, 
$\lambda_a,\ a=0,1,\ldots ,8,$ are normalized such that $\mbox{tr} 
(\lambda_a \lambda_b )=2\delta_{ab}$. Here 
$\lambda_0=\sqrt{\frac{2}{3}}\, 1$, and 
$\lambda_k,\ k=1,2,\ldots ,8$ are the standard $SU(3)$ Gell-Mann 
matrices. It is well-known that such four-quark interactions lead
(for some values of the model parameters) to the formation of a quark
condensate, which is invariant under the vector subgroup $U(3)$ and
thus breaks chiral invariance of the ground state in accordance with
the requirements of three-flavour QCD. 

The 't Hooft determinantal interaction is described by the Lagrangian 
\begin{equation}
\label{Ldet}
  {\cal L}_{6q} =\kappa (\mbox{det}\ \bar{q}P_Lq
                         +\mbox{det}\ \bar{q}P_Rq)
\end{equation}
where the matrices $P_{L,R}=(1\mp\gamma_5)/2$ are chiral projectors 
and the determinant is over f\mbox{}lavour indices. This interaction 
breaks explicitly the axial $U_A(1)$ symmetry, lifting the degeneracy of
$\eta$ and $\eta'$ meson masses (in the chiral limit), and violates 
Zweig's rule \cite{Okubo:1963} due to flavour mixing. It affects also 
the scalar singlet and octet states pushing down the mass of the
$SU(3)$ singlet. 

The large $N_c$ behaviour of the model is reflected in the 
dimensionfull coupling constants, $[G]=M^{-2},\, [\kappa ]=M^{-5}$, 
which count as $G\sim 1/N_c$, $\kappa\sim 1/N_c^{N_f}$. As a result 
the NJL interaction (\ref{L4q}) dominates over ${\cal L}_{6q}$ at 
large $N_c$, as one would expect, because Zweig's rule is exact at 
$N_c=\infty$. These couplings have opposite signs: $G>0,\, \kappa <0$. 

The eight-quark Lagrangian which describes the spin zero interactions 
contains two terms: ${\cal L}_{8q}={\cal L}_{8q}^{(1)} + 
{\cal L}_{8q}^{(2)}$ \cite{Osipov:2005b}, where
\begin{eqnarray}    
   {\cal L}_{8q}^{(1)}&\!\!\! =\!\!\! & 
   8g_1\left[ (\bar q_iP_Rq_m)(\bar q_mP_Lq_i) \right]^2 
   = \frac{g_1}{32}\left[ \mbox{tr} (S-iP)(S+iP)\right]^2
   \nonumber \\
   &\!\!\! =\!\!\!&
   \frac{g_1}{8}\left( S_a^2 + P_a^2\right)^2, \\
   {\cal L}_{8q}^{(2)}&\!\!\! =\!\!\!& 
   16 g_2\left[ (\bar q_iP_Rq_m)(\bar q_mP_Lq_j) 
   (\bar q_jP_Rq_k)(\bar q_kP_Lq_i) \right] 
   \nonumber \\
   &\!\!\! =\!\!\!& 
   \frac{g_2}{16}\mbox{tr}\left[ (S-iP)(S+iP)(S-iP)(S+iP)\right]
   \nonumber \\
   &\!\!\! =\!\!\!& 
   \frac{g_2}{16}\mbox{tr} \left( S^4+P^4+4P^2S^2-2PSPS \right)   
   \nonumber \\
   &\!\!\! =\!\!\!& 
   \frac{g_2}{8}\left[ d_{abe}d_{cde} \left(
   S_aS_bS_cS_d + P_aP_bP_cP_d + 2S_aS_bP_cP_d \right) \right.
   \nonumber \\
   &\!\!\! +\!\!\!& \left.
   4 f_{ace}f_{bde} S_aS_bP_cP_d \right].
\end{eqnarray}
Here the trace is taken over flavour indices $i,j =1,2,3$; the matrices 
$S,\, P$ are given by $S_{ij}=S_a (\lambda_a)_{ij}=2\bar q_jq_i,\
P_{ij}=P_a (\lambda_a)_{ij}=2\bar q_j (i\gamma_5) q_i$. The $f_{abc}$ 
are the well-known totally antisymmetric structure constants: 
$[\lambda_a, \lambda_b]=2i f_{abc}\lambda_c$. The $d_{abc}$ are
totally symmetric quantities: $\{\lambda_a, \lambda_b\} = 2d_{abc}
\lambda_c$. ${\cal L}_{8q}$ is a $U(3)_L\times U(3)_R$ symmetric 
interaction with OZI-violating ef\mbox{}fects in ${\cal L}_{8q}^{(1)}$.
 
The eight-quark interactions ${\cal L}_{8q}$ are the lowest order
terms in number of quark f\mbox{}ields which stabilize the vacuum
state of the model. We restrict our consideration to theses terms, 
because in the long wavelength limit (or in the case when the 
multi-quark correlators create a hierarchy) the higher dimensional 
operators are suppressed.  

Since the coupling constants $G, \kappa, g_1, g_2$ are dimensionful,
the model is not renormalizable. We use the cutof\mbox{}f $\Lambda$ to 
make quark loops f\mbox{}inite. The regularization procedure (Pauli -- 
Villars) is standard and can be found, for instance, in our paper 
\cite{Osipov:2004NPA}, where the regularization function is introduced 
to def\mbox{}ine the coincidence limit of the Schwinger -- DeWitt 
representation for the real part of the quark-loop ef\mbox{}fective 
action. This method is used for our calculations of mass spectra
in the following.

%%%%%%%%%%%%%%%%%%%%%%%%%%%%%%%%%%%%%%%%%%%%%%%%%%%%%%%%%%%%%%%%%%%%%%%%%%
\subsection*{\it 2.2 Stability conditions for the vacuum}
%%%%%%%%%%%%%%%%%%%%%%%%%%%%%%%%%%%%%%%%%%%%%%%%%%%%%%%%%%%%%%%%%%%%%%%%%%
The eight-quark forces stabilize the vacuum \cite{Osipov:2005b}. To 
clarify the meaning of this statement, consider, for the sake of 
simplicity, the effective potential $V(M)$ which one obtains as a
result of bosonization of these multi-quark vertices in the chirally 
symmetric limit $(m=0)$ and in the one-quark-loop approximation 
\begin{eqnarray}
\label{Vh}
   V(M)&\!\!\! =\!\!\!& \frac{h^2}{16}\left( 12G+\kappa h 
             +\frac{27}{2}\lambda h^2\right) \nonumber\\
   &\!\!\! -\!\!\!& \frac{3N_c}{16\pi^2}\left[ M^2J_0(M^2)
             +\Lambda^4 \ln\left(1+\frac{M^2}{\Lambda^2}\right)\right],
\end{eqnarray}
with $\Lambda$ being an ultraviolet cutoff in the quark one-loop 
diagrams, and 
\begin{equation}
\label{j0}
   J_0(M^2)=\Lambda^2- M^2\ln\left(1+\frac{\Lambda^2}{M^2}\right).
\end{equation}
The dependence on the variable $h$ is defined by the stationary phase
equation
\begin{equation}
\label{spa1}
    M+Gh+\frac{\kappa}{16}h^2+\frac{3}{4}\lambda h^3=0,
    \qquad \lambda \equiv g_1+\frac{2}{3}g_2 
\end{equation}
as a function of the model parameters and the argument $M$. 

We start the discussion of the effective potential with the standard
case of four-quark interactions, where the curvature of the potential
at the origin and the sign of the coupling G of the interaction fully 
determine the existence of a globally stable system, which can occur
either in the Wigner-Weyl or in the phase of spontaneously broken
chiral symmetry. Often the word ``instability" is used in connection
with the transition from this symmetric to the spontaneously broken
vacuum at a critical value of $G\Lambda^2$. This is not what is meant 
when we say that the vacuum is unstable. As we hope will be clear
after the discussion presented in the remaining of this section, the 
instability we refer to is an essential pathology of the vacuum,
present in the model with combined four and six quark interactions: we 
show that it is crucial for the stability of the vacuum that the
stationary phase equation ($\ref{spa1}$) possesses only one single
real root when higher order multi-quark interactions are present. We
argue that the enlarged system with six-quark interactions fails in
this respect and that eight-quark interactions are necessary to
stabilize the vacuum. Fig. 1 will illustrate the various stages of the 
discussion. 

%%%%%%%%%%%%%%%%%%%%%%%%%%%    FIG.1    %%%%%%%%%%%%%%%%%%%%%%%%%%%%%%
\begin{figure}[t]
\centerline{\epsfig{file=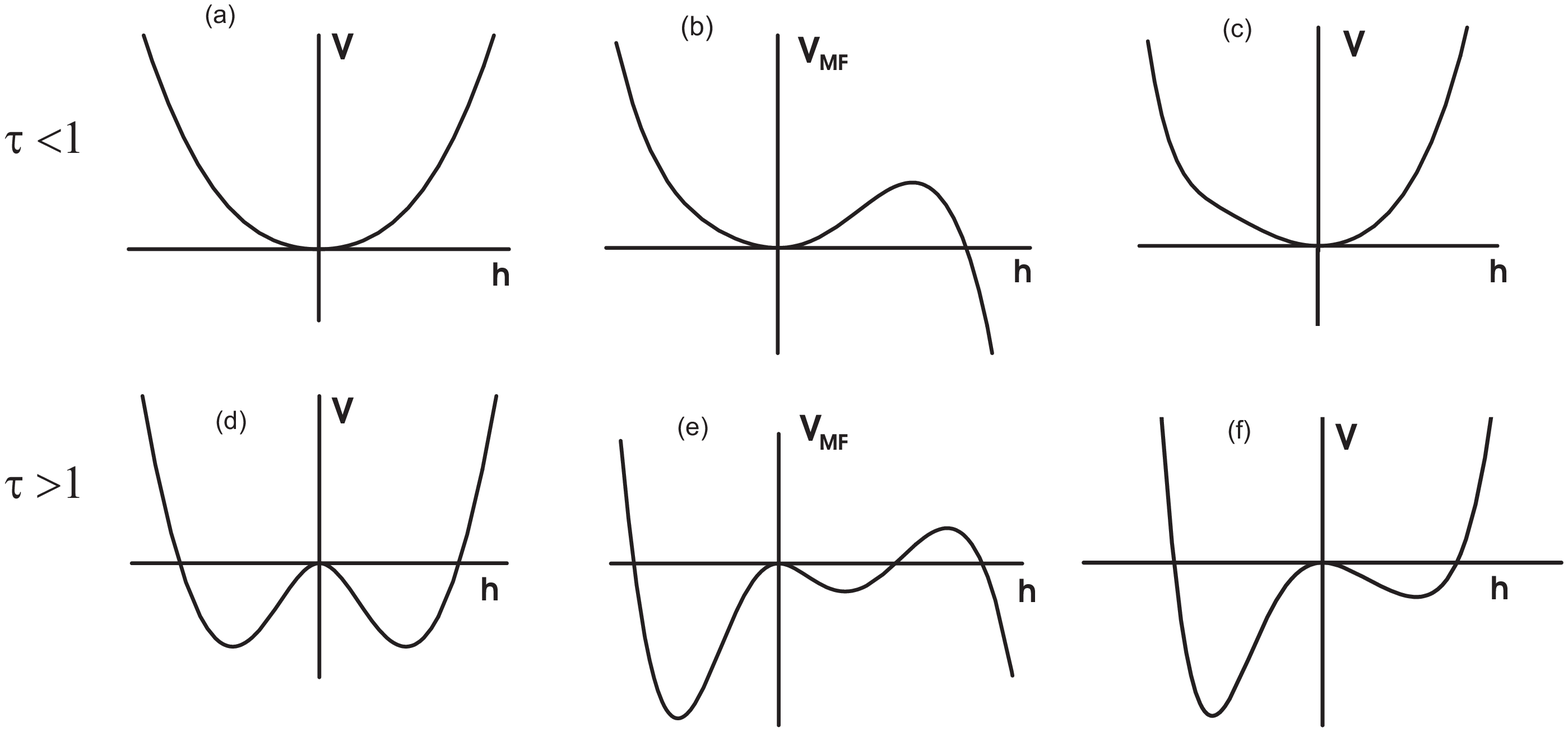,height=7.0cm,width=12cm}}
\caption{\small The effective potential $V$ in the $SU(3)$ chiral
  limit, as function of the quark condensate related variable $h$. 
  Upper and lower panels are classified by the value of 
  $\tau=\frac{N_c G\Lambda^2}{2\pi^2}$, related to the curvature of 
  the effective potential at the origin. Each panel shows the typical 
  form of the potential when one adds successively to the four-quark 
  (left), the six- (middle) and eight-quark interactions (right). 
  The metastable cases (b) and (e) are obtained in the mean field 
  approximation $V_{MF}$ (the stationary phase approach leads to an 
  unstable vacuum, without any local minimum, \cite{Osipov:2005a}).} 
\label{fig1}
\end{figure}
%%%%%%%%%%%%%%%%%%%%%%%%%%%%%%%%%%%%%%%%%%%%%%%%%%%%%%%%%%%%%%%%%%%%%%%%  

In a world without six and eight-quark interactions, $\kappa, g_1, 
g_2=0$, one obtains from Eq. (\ref{spa1}) that $h=-M/G$ and, as a 
result, the potential $V(M)$ has the form of a double well, if 
$V''(0)<0$, see Fig. 1d, i.e. if $\tau$, the following combination of 
model parameters,  
\begin{equation}
\label{ssb}
  \tau= \frac{N_c G\Lambda^2}{2\pi^2}>1.
\end{equation}   
This inequality expresses the fact that chiral symmetry is
spontaneously broken{\footnote {The Wigner-Weyl phase appears for 
$0<\tau<1$, i.e. $V''(0)>0$, see Fig. 1a.}}, producing in the massless 
case, $(m=0)$, the degeneracy of a nonet of Goldstone bosons, and 
showing the presence of the $U_A(1)$ problem. This vacuum state is 
globally stable, because at large values of $|M|$ another inequality
is fulfilled
\begin{equation}
\label{asympt}
   V(M) \sim \frac{3G}{4} h^2(M) = 
   \frac{3M^2}{4G} >0 \qquad (M\to\pm\infty ) 
\end{equation}
at $G>0$. In this particular case the stability of the
ground state $(G>0)$ is already guaranteed by Eq. (\ref{ssb}).  

It is worth noting that higher order multi-quark interactions will
not change condition (\ref{ssb}) as long as Eq. (\ref{spa1}) has only 
one real solution.  The reason for this is very simple. If this
equation has only one real solution, it is valid to expect that 
$h(M)=-M/G + {\cal O}(M^2)$. Since $2n$-quark vertices contribute to
$V(M)$ as $h^n\sim M^{n}$, the value of $V''(0)$ is entirely
determined by terms of the second power in $M$, i.e. by the four-quark
interaction $(n=2)$ only. Therefore it is tempting to describe the
general situation (when higher order multi-quark interactions are 
included) by the same inequality (\ref{ssb}), since it will dictate
the behavior of the effective potential in the neighbourhood of the
origin in full agreement with the leading order result; this is 
illustrated in Fig. 1, the upper panel for $\tau<1$, the lower one for 
$\tau>1$. For that one must find however a way to reduce the number of 
real roots of the corresponding stationary phase equation to one. 

What is wrong with several roots? Let us consider the system which
includes four- and six-quark interactions $G,\kappa \neq 0,\, g_1,
g_2=0$. In this case the quadratic Eq. (\ref{spa1}) has two solutions, 
both being real for $M\geq 4G^2/\kappa$. It follows then that the 
stationary phase method leads us to the gap equation which contains
the sum of these solutions
\begin{equation}
\label{gap2}
   h^{(1)}+h^{(2)}+\frac{N_cM}{\pi^2}J_0(M^2)=0.
\end{equation}
The sum does not depend on $M$, because $h^{(1)}+h^{(2)}=-16G/\kappa$, 
and this gap equation misses the trivial solution $M=0$, corresponding
to the chiral symmetric vacuum. One sees that a simple addition of the 
't Hooft interaction to the four-quark Lagrangian affects so violently 
the trivial solution and as a matter of fact the whole effective
potential, which gets unstable\footnote{This point has been considered 
in detail in \cite{Osipov:2005a}.}, that, apparently, we must get rid
of the problem which appears as soon as the stationary phase equation
has more than one real solution\footnote{ 
Here we would like to stress that Figs. 1b and 1e, related with the 
addition of six-quark interactions, are obtained within the mean field
approach, which leads to the effective potential (\ref{Vh}) taken at 
$\lambda=0$ and considered as a function of $h$ \cite{Hatsuda:1994}. 
We identify it with $V_{MF}(h)$. The dependence $M(h)$ is
given by Eq. (\ref{spa1}). This is a one-to-one mapping $h\to M$, where 
$h$ ranges along the interval $-\infty <h<\infty$. The local maximum
at positive $h=-8G/ \kappa$ on both figures corresponds to the point 
where the regular (at $\kappa\to 0$) solution $h^{(1)}$ changes to the 
singular $h^{(2)}$. $V_{MF}(h)$ is unbounded from below, as 
$h^{(2)}\to\infty$.}.

Since a quadratic equation never has only one real root, we are pushed
to increase the order of the equation by including eight-quark
interactions. One obtains in this way the cubic equation (\ref{spa1}). 
This equation has only one real solution $h(M)$, which changes
smoothly in the open interval $-\infty <h<\infty $ being an isomorphic 
and monotonic function of $M$, when one restricts the choice of 
parameters to 
\begin{equation}
\label{stabcond}
   G>\frac{1}{\lambda}\left( \frac{\kappa}{24}\right)^2, 
   \qquad \lambda >0.
\end{equation}
For this case the constituent quark mass $M$ fulfills the gap equation
\begin{equation}
\label{gap1}
   h(M) + \frac{N_cM}{2\pi^2}J_0(M^2)=0,
\end{equation}
related with the potential $V(M)$ (see Eq. (\ref{Vh})) which is bounded 
from below (see Figs. 1c and 1f).   

Eqs. (\ref{stabcond}) replace the previous requirement $G>0$ (see 
Eq. (\ref{asympt})). They must be fulfilled to guarantee the global 
stability of the system. The first inequality is new and plays for the 
enlarged system the role of Eq. (\ref{asympt}): in the case of
four-quark interactions only, the linear stationary phase equation 
had automatically only one real root, here the values of couplings 
must be fixed correspondingly to ensure the existence of only one real 
root. The second inequality is a direct analogue of $G>0$.  

Eq. (\ref{ssb}) is still relevant to the case and is responsible for
the behaviour of $V(M)$ in the neighbourhood of zero, as mentioned
before. In Figs. 1c and 1f the stabilizing effect due to the addition
of eight-quark interactions is shown. Note that they change radically 
the potential only at values of $h> -8G/\kappa$, as compared to the 
cases 1b and 1e, calculated in the mean field approximation (see also
footnote 7), affecting little the other branch of the potential, where
$h\sim h^{(1)}$. In particular the value of $h$ where the global
minimum of both potentials occurs in the spontaneously broken phase, is 
negative. Since at the quark one-loop order $h$ is proportional to the 
quark condensate \cite{Osipov:2004NPA}, one is inclined to believe 
that by fixing the model parameters through it, this will finally lead to 
similar numerical values for all observables which depend in a 
stringent way on the value of the condensate. For those observables, the 
calculatinos in the metastable mean-field approximation and in the 
globally stable case considered with inclusion of the eight-quark 
interactions will not differ much. 

%%%%%%%%%%%%%%%%%%%%%%%%%%%    FIG.2    %%%%%%%%%%%%%%%%%%%%%%%%%%%%%%
\begin{figure}[t]
\centerline{\epsfig{file=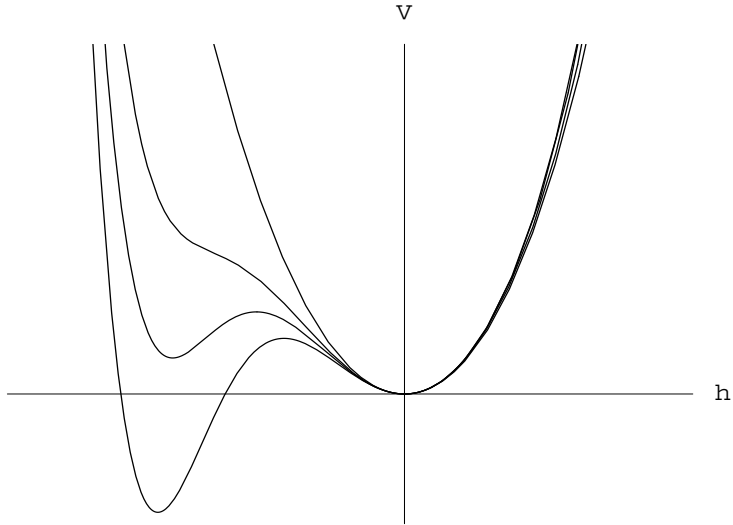,height=7.0cm,width=10cm}}
\caption{\small A closer view on the effective potential $V$ of
  Fig. 1c. Depending on the strength of the six-quark coupling
  $\kappa$, with remaining parameters $G,\lambda,\Lambda$ fixed, the 
  symmetric Wigner-Weyl phase of the four-quark potential (Fig. 1a) may 
  coexist with the spontaneously broken phase induced by the presence of 
  the 't Hooft term. This double vacuum exists within the global 
  stability conditions (\ref{stabcond}).} 
\label{fig2}
\end{figure}
%%%%%%%%%%%%%%%%%%%%%%%%%%%%%%%%%%%%%%%%%%%%%%%%%%%%%%%%%%%%%%%%%%%%%%%%  

Another interesting aspect of this simple analysis of the $SU(3)$
limit of the effective potential is the possibility of existence of 
multiple vacua, illustrated in Fig. 2. For $\tau<1$, i.e. in a region 
where the four-quark interactions alone lead to the symmetric
Wigner-Weyl phase, the inclusion of the 't Hooft six-quark
intearctions can induce spontaneous symmetry breaking for some
critical values of the coupling parameter $\kappa$. This new vacuum
can coexist with the trivial vacuum. This has been discussed
previously in \cite{Osipov:2004}, now we are able to confirm its 
existence within the general stability conditions imposed for the 
vacuum, given by the inequalities in (\ref{stabcond}). In the presence 
of non-vanishing values for the current quark mass, the minimum at the 
origin shifts towards the physical region of negative $h$. The
phenomenon of multiple vacua has been addressed in several other 
approaches to the description of the QCD vacuum 
\cite{Veneziano:1980,Halperin:1998,Bicudo:2006}.          

The most general case is obtained if the $SU(3)_L\times SU(3)_R$
chiral symmetry of the quark Lagrangian is broken down explicitly by
the non-zero values of current quark masses $m$. Then the inequalities 
(\ref{stabcond}) must be replaced by the following ones 
\cite{Osipov:2005b}
\begin{equation}
\label{ineq1}
   g_1>0, \quad g_1 +3g_2>0, \quad 
   G>\frac{1}{g_1}\left(\frac{\kappa}{16}\right)^2.
\end{equation}  

The last constraint can be used to make a large $N_c$ estimate for 
$g_1$. Indeed, we know that $G$ scales as $1/N_c$, $\kappa\sim
1/N_c^3$ and, therefore, conclude from the above inequality that 
$g_1$ cannot scale as $1/N_c^6$ or smaller. On the other hand, this 
eight-quark interaction is an additional (to the 't Hooft
determinant) source of OZI-violating ef\mbox{}fects and thus it 
cannot be stronger than the 't Hooft interactions, i.e., $g_1\sim 1/N_c^4$ 
or less. These reasonings show that $1/N_c^5 \leq g_1 \leq 1/N_c^4$. 
We would expect $g_2$ to be in the same interval, although
inequalities (\ref{ineq1}) are not appropriate to prove that.

%%%%%%%%%%%%%%%%%%%%%%%%%%%%%%%%%%%%%%%%%%%%%%%%%%%%%%%%%%%%%%%%%%%%%%%%%%
\subsection*{\it 2.3 Semi-bosonized Lagrangian}
%%%%%%%%%%%%%%%%%%%%%%%%%%%%%%%%%%%%%%%%%%%%%%%%%%%%%%%%%%%%%%%%%%%%%%%%%%
The multi-quark Lagrangian (\ref{efflag}) can be presented in the
bilinear form with respect to the quark fields. The details can be 
found in \cite{Reinhardt:1988, Osipov:2005a}. Without using any 
approximations one obtains the following vacuum-to-vacuum amplitude 
of the theory
\begin{eqnarray}
\label{genf}
   Z&\!\!\! =\!\!\!&\int {\cal D}q{\cal D}\bar{q}
     \prod_a{\cal D}\sigma_a\prod_a{\cal D}\phi_a\ 
     \exp\left(i\int\ud^4x
     {\cal L}_q(\bar{q},q,\sigma ,\phi )\right)
     \nonumber \\
    &\!\!\!\times\!\!\!& \int\limits^{+\infty}_{-\infty}
     \prod_a{\cal D}s_a\prod_a{\cal D}p_a\
     \exp\left(i\int\ud^4x{\cal L}_r(\sigma ,\phi ,\Delta ;s,p)\right),
\end{eqnarray}  
where
\begin{eqnarray}
\label{lagr2}
  {\cal L}_q &\!\!\! =\!\!\!&
  \bar{q}(i\gamma^\mu\partial_\mu -M-\sigma - i\gamma_5\phi )q, \\
\label{Lr}
   {\cal L}_r &\!\!\! =\!\!\!& s_a (\sigma_a+\Delta_a) + p_a\phi_a
   +\frac{G}{2} \left(s_a^2+p_a^2 \right) 
   \nonumber \\
   &\!\!\! +\!\!\!& \frac{\kappa}{32}\ A_{abc}s_a
   \left(s_bs_c-3p_bp_c\right)
   + \frac{g_1}{8} \left(s_a^2+p_a^2\right)^2
   \nonumber \\
   &\!\!\! +\!\!\!& \frac{g_2}{8}\left[d_{abe}d_{cde}\left(
   s_as_bs_cs_d + 2s_as_bp_cp_d +p_ap_bp_cp_d \right)\right.
   \nonumber \\
   &\!\!\! +\!\!\!& \left. 4f_{ace}f_{bde} s_as_bp_cp_d\right]. 
\end{eqnarray}

The bosonic f\mbox{}ields $\sigma =\sigma_a\lambda_a$ and $\phi 
=\phi_a\lambda_a$ are the composite scalar and pseudoscalar nonets
which will be identif\mbox{}ied later with the corresponding 
physical states. The auxiliary f\mbox{}ields $s_a$ and $p_a$ must be 
integrated out from the ef\mbox{}fective mesonic Lagrangian 
${\cal L}_r$. The quarks obtain their constituent masses 
$M=M_a\lambda_a=\mbox{diag}(M_u,M_d,M_s)$ due to dynamical chiral 
symmetry breaking in the physical vacuum state, 
$\Delta_a=M_a - m_a$. The totally symmetric constants $A_{abc}$
are related to the f\mbox{}lavour determinant, and equal to
\begin{equation}
\label{A}
   A_{abc}=\frac{1}{3!}\epsilon_{ijk}\epsilon_{mnl}(\lambda_a)_{im}
             (\lambda_b)_{jn}(\lambda_c)_{kl}. 
\end{equation}
Some useful relations for $A_{abc}$ can be found in 
\cite{Osipov:2004}.

%%%%%%%%%%%%%%%%%%%%%%%%%%%%%%%%%%%%%%%%%%%%%%%%%%%%%%%%%%%%%%%%%%%%%%%%%%
\subsection*{\it 2.4 Stationary phase approximation for $Z$}
%%%%%%%%%%%%%%%%%%%%%%%%%%%%%%%%%%%%%%%%%%%%%%%%%%%%%%%%%%%%%%%%%%%%%%%%%%
The functional integrals over auxiliary variables $s_a, p_a$ in 
Eq. (\ref{genf}) can be calculated approximately within the stationary 
phase method. For that one should f\mbox{}irst f\mbox{}ind all real 
stationary phase trajectories $s_a^{st}=s_a(\sigma, \phi ),\ p_a^{st}
= p_a(\sigma, \phi )$ given by the system of equations   
\begin{equation} 
\label{fdLr}
   \frac{\partial {\cal L}_r}{\partial s_a} = 0, \qquad 
   \frac{\partial {\cal L}_r}{\partial p_a} = 0. 
\end{equation}

We seek these solutions in form of expansions in the external mesonic 
f\mbox{}ields $\sigma_a , \phi_a$ 
\begin{eqnarray}
\label{Rst}
   s_a^{st} &\!\!\! =\!\!\!& h_a + h_{ab}^{(1)}\sigma_b  
            + h_{abc}^{(1)}\sigma_b\sigma_c 
            + h_{abc}^{(2)}\phi_b\phi_c + \ldots 
            \nonumber \\
   p_a^{st} &\!\!\! =\!\!\!& h_{ab}^{(2)}\phi_b 
            + h_{abc}^{(3)}\phi_b\sigma_c + \ldots 
\end{eqnarray}
The coef\mbox{}f\mbox{}icients $h_{a\ldots}^{(i)}$ depend on the 
coupling constants $G,\kappa, g_1, g_2$ and quark masses $\Delta_a$. 
The higher index coef\mbox{}f\mbox{}icients $h_{a\ldots}^{(i)}$ are 
recurrently expressed in terms of the lower ones. The one-index 
coef\mbox{}f\mbox{}icients $h_a$ are the solutions of the following 
system of cubic equations 
\begin{equation}
\label{res}
   \Delta_a + Gh_a + \frac{3\kappa}{32} A_{abc} h_bh_c 
   +\frac{g_1}{2} h_a h_b^2 +\frac{g_2}{2} d_{abe}d_{cde} h_bh_ch_d=0.
\end{equation}
Thus, the problem is reduced to a finite set of algebraic equations
which span the nonet space of $U(3)$. These equations may be 
considered as an example of response equations, previously studied in
\cite{Pais:1968} for the purpose of gaining some understanding about
the possible origins of strong $SU(3)$ breaking. The response $h_a$
is fully determined by the couplings $G, \kappa, g_1, g_2$ and the mean 
f\mbox{}ield $\Delta_a$, which plays the role of an external force. In 
accordance with the pattern of explicit symmetry breaking the mean 
f\mbox{}ield has only three non-zero components, with indices 
$a=0,3,8$. 

In order to solve Eqs. (\ref{res}), one has to find whether there exists
an intersection of a number of hypersurfaces. The important question of 
completeness of the system (\ref{res}) is examined in Appendix A. This 
yields $h_a=0$ for $a=1,2,4,5,6,7$. Thus the system reduces to three 
coupled equations to determine $h_0, h_3, h_8$. This task has been solved
in \cite{Osipov:2005b}. At this stage, as we have already discussed, one
has to find conditions (see Eq. (\ref{ineq1})) which ensure the
one-to-one mapping $\Delta_a\leftrightarrow h_a$.  

The next two equations following from (\ref{fdLr}) determine the 
coefficients $h_{ar}^{(1)},\, h_{ar}^{(2)}$ in (\ref{Rst}) 
\begin{eqnarray}
\label{hab1}
   &&\left\{\left(G+\frac{g_1}{2} h_b^2\right)\delta_{ar} 
     +\frac{3\kappa}{16} A_{abr} h_b + g_1 h_a h_r 
     \right.\nonumber \\
   &&\left.+\frac{g_2}{2} \left(2 d_{abe} d_{rce}  
     + d_{cbe} d_{rae}\right) h_c h_b\right\}h_{rs}^{(1)}
     = -\delta_{as},
\end{eqnarray}
\begin{eqnarray}
\label{hab2}
   &&\left\{\left(G+\frac{g_1}{2}\, h_b^2\right)\delta_{ar} 
     -\frac{3\kappa}{16} A_{abr} h_b 
     \right.\nonumber \\
   &&+\left.\frac{g_2}{2}\left( 2f_{abe} f_{rce}
     + d_{cbe} d_{rae}\right) 
     h_c h_b\right\} h_{rs}^{(2)}=-\delta_{as}.
\end{eqnarray}
Corresponding solutions are given in Appendix B. 

This procedure can be easily extended. Equating to zero the factor at
any independent field combination in (\ref{fdLr}), one obtains an
equation which determines one of the coefficients in (\ref{Rst}). 

On the other hand, these equations are useful if one wants to find the 
projection of the Lagrangian ${\cal L}_r$ on the stationary phase 
trajectory (\ref{Rst}). For that one should rewrite them in a more
convenient form, using that
\begin{eqnarray}
  (h^{(1)})_{ar}^{-1} h_r
  &\!\!\! =\!\!\!&Gh_a + 2 \Delta_a
   -\frac{g_1}{2} h_a h_b^2 
   - \frac{g_2}{2} d_{abe} d_{rce}h_b h_c h_r\, , \\
   -(h^{(2)})_{ar}^{-1} h_r
  &\!\!\! =\!\!\!&3Gh_a+2 \Delta_a + 3\frac{g_1}{2}h_a h_b^2
   +3\frac{g_2}{2} d_{abe} d_{cde}h_b h_c h_d
   \nonumber \\
  &\!\!\! +\!\!\!& g_2 f_{ace} f_{rde} h_c h_d h_r\, .
\end{eqnarray}

In particular, solutions of Eqs. (\ref{res})-(\ref{hab2}) define the
first three coupling constants of such Lagrangian, i.e., one can show 
that
\begin{equation}
\label{lr}
   {\cal L}_r\to
   {\cal L}_{st}= h_a \sigma_a + 
   \frac{1}{2}\,h_{ab}^{(1)} \sigma_a\sigma_b 
   + \frac{1}{2}\, h_{ab}^{(2)}\,\phi_a\phi_b 
   + {\cal O}(\mbox{field}^3).
\end{equation}

Since the system of equations (\ref{fdLr}) can be solved, we are able 
to obtain the semi-classical asymptotics of the functional integral
over $s_a$ and $p_a$ in $Z$. If parameters $G, \kappa, g_1, g_2$ 
belong to a range where the system has a unique real solution, the 
calculations are straightforward. In particular, one has the following 
result which is valid at lowest order of the stationary phase 
approximation
\begin{eqnarray}
\label{intJisp}
   &&\int\limits^{+\infty}_{-\infty}\prod_a{\cal D}s_a
     \prod_a{\cal D}p_a\ \exp\left(i\int\ud^4x{\cal L}_r
     (\sigma ,\phi ,\Delta ;s,p)\right) \nonumber \\
   &&\sim\ \exp\left(i\int\ud^4x{\cal L}_{st}(\sigma ,
     \phi )\right) 
     \qquad (\hbar\to 0).
\end{eqnarray}

%%%%%%%%%%%%%%%%%%%%%%%%%%%%%%%%%%%%%%%%%%%%%%%%%%%%%%%%%%%%%%%%%%%%%%%%%%
\subsection*{\it 2.5 Integrating quark fields in $Z$}
%%%%%%%%%%%%%%%%%%%%%%%%%%%%%%%%%%%%%%%%%%%%%%%%%%%%%%%%%%%%%%%%%%%%%%%%%%
To obtain the effective Lagrangian of the model we should integrate 
out quark fields from Eq. (\ref{genf}). This is a well studied part of the
calculations and we restrict ourselves to several general remarks here.
 
The one-quark-loop effective action can be computed in euclidean 
spacetime, the chiral invariant part of the result, $W_q[\sigma, \phi
]$, is given by the modulus of the quark determinant 
\begin{equation}
   W_q[\sigma, \phi ]=\ln |\det D_E|,
\end{equation}
where $D_E$ stands for the Dirac operator in euclidean spacetime,
namely $D_E=i\gamma_\mu\partial_\mu -M-\sigma -i\gamma_5\phi$.
The quark determinant is a complicated nonlocal functional which can
be approximated in the low-energy regime by the Schwinger -- DeWitt 
asymptotic expansion \cite{Schwinger:1951,Ball:1989}. The presence of
a noncommutative (with respect to the bosonic fields $\sigma$ and
$\phi$) mass matrix $M$ requires a more delicate treatment of this term 
in comparison with the standard approach, where $M$ is supposed to 
commutative with the fields. The corresponding technique has been recently
developed \cite{Osipov:2001} and applied to the case considered here 
in \cite{Osipov:2004NPA}. We refer to these papers for necessary
details (see Section 3 in \cite{Osipov:2004NPA}), although we 
present the result, because we need it in the following. 

The heat kernel expansion used is
\begin{equation}
\label{hkexp}
   W_q[\sigma, \phi ]=-\!\int\!\frac{\ud^4x_E}{32\pi^2}
   \sum_{i=1}^{\infty} I_{i-1}\mbox{tr}(b_i),
\end{equation}
where coefficients $b_i$ for the case with isospin symmetry are
\begin{equation}
\label{b1b2}
   b_1=-Y, \qquad b_2=\frac{Y^2}{2} + \frac{\Delta_{us}}{\sqrt{3}}
   \lambda_8 Y, \qquad \ldots
\end{equation}
Our following result is based on these two terms of the series. It is
the lowest order approximation, because the usual kinetic term of the 
collective fields is contained in $b_2$, and we truncate the series 
exactly after this term. The part of $b_2$ with $\Delta_{us}=M_u^2-M_s^2$ 
is absent from the standard Seeley -- DeWitt coefficient 
$a_2$. This is one of the new features of the approach, which follows from
the noncommutativity of the constituent quark mass matrix $M$. 

The trace in Eq. (\ref{hkexp}) should be taken over colour, flavour and 
four-spinors indices. In Eq. (\ref{b1b2}) $Y$ is used for 
\begin{equation}
   Y=i\gamma_\mu (\partial_\mu\sigma +i\gamma_5\partial_\mu\phi )
    +\sigma^2 + [M,\sigma ] + \phi^2 +i\gamma_5[\sigma +M,\phi].
\end{equation} 
The factors $I_i$ are given by the average
\begin{equation}
   I_i =\frac{1}{3}\left[2J_i(M_u^2)+J_i(M_s^2)\right].
\end{equation}
and represent one-quark-loop integrals. In the 
considered approximation we need only to know $J_0(M^2)$ (see 
Eq. (\ref{j0})) and
\begin{equation}
      J_1(M^2)=\ln\left(1+\frac{\Lambda^2}{M^2}\right) 
      -\frac{\Lambda^2}{\Lambda^2+M^2}\, .
\end{equation}
We are using here the proper time regularization scheme.

Thus, the integration over quark fields yields the second part of the
effective Lagrangian (the first part is given by Eq. (\ref{lr}))
\begin{equation}
   {\cal L}_q \to {\cal L}_{hk} = 
   {\cal L}_{tad} + {\cal L}_{kin} + {\cal L}_{m} 
   + {\cal L}_{int}. 
\end{equation}
 
The tadpole term, ${\cal L}_{tad}$, is
\begin{equation} 
   {\cal L}_{tad}=\frac{N_c}{12\pi^2}\left[ 
   M_u (3I_0-\Delta_{us}I_1)(\sigma_u +\sigma_d ) 
   + M_s (3I_0 +2\Delta_{us}I_1) \sigma_s \right]. 
\end{equation}

The kinetic term, ${\cal L}_{kin}$, after continuation to the
Minkowski spacetime, requires a redefinition of meson fields  
to obtain the standard factor in front, i.e.,
\begin{equation}
\label{kin}
   {\cal L}_{kin}=\frac{N_cI_1}{16\pi^2}\mbox{tr}\left[ 
   (\partial_\mu\sigma )^2+(\partial_\mu\phi )^2\right] 
   = \frac{1}{4}\mbox{tr}\left[ 
   (\partial_\mu\sigma_R )^2+(\partial_\mu\phi_R )^2\right], 
\end{equation}
where 
\begin{equation}
   \sigma^a = g\sigma^a_R, \quad 
   \phi^a = g\phi^a_R, \quad g^2=\frac{4\pi^2}{N_cI_1}\, .
\label{g}
\end{equation}

The contribution to the mass Lagrangian is given by
\begin{eqnarray}
\label{mass}
     {\cal L}_{m}
     \!\!\!\!\!\!\!\!\!
     &&=\frac{N_cI_0}{4\pi^2}\left( \sigma_a^2+\phi_a^2\right)  
        -\frac{N_cI_1}{12\pi^2}
        \left\{
        \Delta_{us}[2\sqrt{2}(3\sigma_0\sigma_8+\phi_0\phi_8)
        -\phi_8^2+\phi_i^2]
        \right.\nonumber \\
     &&\ \ \ +2(2M_u^2+M_s^2)\sigma_0^2+(M_u^2+5M_s^2)\sigma_8^2
       +(7M_u^2-M_s^2)\sigma_i^2 \nonumber \\
     &&\ \ \ \left. +(M_u+M_s)(M_u+2M_s)\sigma_f^2
       +(M_s-M_u)(2M_s-M_u)\phi_f^2\right\},
\end{eqnarray}
where we assume that the indices $i$ and $f$ range over the subsets
$i=1,2,3$ and $f=4,5,6,7$ of the set $a=0,1,\ldots,8.$ Thus we have
\begin{eqnarray}
\label{not}
   && \phi_i^2=2\pi^+\pi^-+(\pi^0)^2,\qquad 
      \phi_f^2 = 2(K^+K^-+\bar{K}^0K^0), \nonumber \\
   && \sigma_i^2=2a_0^+a_0^-+(a_0^0)^2,\qquad 
   \sigma_f^2 = 2(K^{*+}_0 K^{*-}_0 + \bar{K}^{*0}_0 
   K^{*0}_0). 
\end{eqnarray}

%%%%%%%%%%%%%%%%%%%%%%%%%%%%%%%%%%%%%%%%%%%%%%%%%%%%%%%%%%%%%%%%%%%%%%%%%%%%%%

\section*{\centerline{\large\bf 3. Pseudoscalars: masses, mixings and
      all that}}

%%%%%%%%%%%%%%%%%%%%%%%%%%%%%%%%%%%%%%%%%%%%%%%%%%%%%%%%%%%%%%%%%%%%%%%%%%%%%%

%%%%%%%%%%%%%%%%%%%%%%%%%%%%%%%%%%%%%%%%%%%%%%%%%%%%%%%%%%%%%%%%%%%%%%%%%%
\subsection*{\it 3.1 Symmetry and currents}
%%%%%%%%%%%%%%%%%%%%%%%%%%%%%%%%%%%%%%%%%%%%%%%%%%%%%%%%%%%%%%%%%%%%%%%%%%
One can simply obtain the conserved (or partially conserved) currents
of the local theory by using the variational method of Gell-Mann and 
L\'{e}vy \cite{Gell-Mann:1960}. For instance, the infinitesimal local
chiral transformations of the quark fields in ${\cal L}_{eff}$ (see 
Eq. (\ref{efflag})) are
\begin{equation}
\label{chiral}
   \delta q      =i(\alpha +\gamma_5\beta )q, \qquad
   \delta\bar{q} =-i\bar{q}(\alpha -\gamma_5\beta ),
\end{equation}
where the small parameters $\alpha =\alpha_a\frac{\lambda_a}{2}$ and
$\beta =\beta_a\frac{\lambda_a}{2}$ are Hermitian flavour matrices.
Then, according to the Gell-Mann -- L\'{e}vy formula, one obtains
the standard vector $V_\mu^a$ and axial-vector $A_{\mu}^a$ 
nonet quark currents
\begin{equation}
   V_\mu^a=-\frac{\delta {\cal L}_{eff}}{\delta (\partial 
   \alpha_a)} = \bar{q}\gamma_\mu\frac{\lambda_a}{2}q, \qquad
   A_\mu^a=-\frac{\delta {\cal L}_{eff}}{\delta (\partial \beta_a)}
   =\bar{q}\gamma_\mu\gamma_5\frac{\lambda_a}{2}q, 
\end{equation} 
and their divergences 
\begin{eqnarray}
     \partial^\mu V_\mu^a = -\frac{\delta {\cal L}_{eff}}{\delta  
     \alpha_a} &\!\!\! =\!\!\!& 
     \frac{i}{2}\, \bar{q}\left[m,\lambda_a\right]q, \\
     \partial^\mu A_\mu^a = -\frac{\delta {\cal L}_{eff}}{\delta  
     \beta_a} &\!\!\! =\!\!\!& 
     \frac{i}{2} \bar{q}\gamma_5\left\{m,\lambda_a \right\}q 
     \nonumber \\
     &\!\!\! +\!\!\!& 
     i\delta_{a0}\sqrt{6}\kappa 
     \left(\det\bar{q}P_Lq - \det\bar{q}P_Rq\right).
\end{eqnarray}

Transformations (\ref{chiral}) induce the correlated change in the flavour 
space of collective fields  
\begin{eqnarray}
\label{inf3}
   \delta \sigma_R\!\!\!\!\!\!\!\!\!
   &&=i\left[\alpha,\sigma_R +Mg^{-1}\right]
      +\left\{\beta,\phi_R\right\}, 
      \nonumber \\
   \delta \phi_R\!\!\!\!\!\!\!\!\!
   &&=i\left[\alpha,\phi_R\right]
      -\left\{\beta,\sigma_R +Mg^{-1}\right\}.   
\end{eqnarray}
The quark Lagrangian, ${\cal L}_{eff}$, is approximated by
the effective bosonized Lagrangian ${\cal L}_{bos}$
\begin{equation}
\label{Lbos}
   {\cal L}_{eff}\to {\cal L}_{bos} = {\cal L}_{st} + {\cal L}_{hk}
                                    + \ldots ,  
\end{equation}
where dots correspond to all omitted terms due to the approximations made.
Therefore, one can obtain the currents again, but now they will be 
written in terms of meson fields. Indeed, one has 
\begin{equation}
\label{axicur}
   {\cal A}_{\mu}^a=\frac{1}{4}\mbox{tr}\left[\left(
              \left\{\sigma_R + Mg^{-1},
              \partial_\mu\phi_R \right\} 
              -\left\{\partial_\mu\sigma_R,\phi_R 
              \right\}\right)\lambda_a\right]+{\cal O}(b_3),   
\end{equation}   
\begin{equation}
\label{veccur}
   {\cal V}_{\mu}^a=-\frac{i}{4}\mbox{tr}\left[\left(
              \left[\sigma_R + Mg^{-1},
              \partial_\mu\sigma_R\right] 
              +\left[\phi_R,\partial_\mu\phi_R
              \right]\right)\lambda_a\right]+{\cal O}(b_3).   
\end{equation}   
These currents obviously depend on the order where the 
heat kernel series is truncated, because the coefficient $b_2$ and 
higher ones contain derivatives. The symbol ${\cal O}(b_3)$ shows 
that currents (\ref{axicur}) and (\ref{veccur}) have been obtained 
from the Lagrangian ${\cal L}_{hk}$ based on two terms of the 
asymptotic series (\ref{hkexp}), namely $b_1$ and $b_2$.

%%%%%%%%%%%%%%%%%%%%%%%%%%%%%%%%%%%%%%%%%%%%%%%%%%%%%%%%%%%%%%%%%%%%%%%%%%
\subsection*{\it 3.2 Decay constants of pseudoscalars}
%%%%%%%%%%%%%%%%%%%%%%%%%%%%%%%%%%%%%%%%%%%%%%%%%%%%%%%%%%%%%%%%%%%%%%%%%%
Let us calculate matrix elements of axial-vector currents
\begin{equation}
   \langle 0|{\cal A}^a_\mu (0)| \phi^b_R (p)\rangle =i 
   f^{ab}p_\mu . 
\end{equation}
Using Eq. (\ref{axicur}) one derives
\begin{eqnarray} 
   && f^{00}=\frac{2M_u+M_s}{3g}\, , \quad
      f^{11}=f^{22}=f^{33}=\frac{M_u}{g}\, , \nonumber \\
   && f^{44}=f^{55}=f^{66}=f^{77}=\frac{M_u+M_s}{2g} 
      \, , \nonumber \\
   && f^{88} = \frac{M_u+2M_s}{3g}\, , \quad
      f^{08} = f^{80} = \sqrt{2}\, \frac{M_u-M_s}{3g}\, .  
\end{eqnarray}

It is not difficult to relate these abstract values with the 
experimentally measured decay constants of physical pseudoscalar 
states $P(x)$. For instance, the weak decay constants of the pion 
($f_\pi$) and kaon ($f_K$) are defined by the corresponding isotopic 
components of the axial current ${\cal A}_{\mu}^{1+i2}$ 
and ${\cal A}_{\mu}^{4+i5}$, i.e., 
\begin{equation}
   \langle 0|{\cal A}^{1+i2}_\mu (0)|\pi (p)\rangle 
   = i \sqrt{2} f_\pi p_\mu , \quad  
   \langle 0|{\cal A}^{4+i5}_\mu (0)|K(p)\rangle 
   =i\sqrt{2} f_K p_\mu , 
\end{equation}
and, therefore, one finds\footnote{We use the normalizations $f_\pi 
=92.42\pm 0.26$ MeV, and $f_K=113.00\pm 1.03$ MeV \cite{PDG:2004}.}
\begin{equation}
\label{GT}
   f_\pi =\frac{M_u}{g}\ , \qquad f_K=\frac{M_s+M_u}{2g}\ .
\end{equation}
These are the Goldberger -- Treiman relations at the quark level. 

The decay constants in the $\eta -\eta'$ system are defined 
\begin{equation}
\label{etamat}
   \langle 0|{\cal A}^a_\mu (0)| P(p)\rangle =i f^{a}_P p_\mu , 
   \quad (a=0,8)
\end{equation}
where $P=\eta, \eta'$. Each of the two mesons has both, singlet and 
octet components
\begin{equation}
\label{singoct}
          { \phi_R^0 \choose \phi_R^8 } = \left( 
          \begin{array}{cc}
          \cos\theta_p & -\sin\theta_p \\
          \sin\theta_p & \cos\theta_p 
          \end{array}
    \right)
          { \eta' \choose \eta }\, . 
\end{equation}
This orthogonal rotation diagonalizes the kinetic and mass terms in the 
meson effective Lagrangian. The mixing angle $\theta_p$ will be 
calculated in Section 3.4. Consequently, from Eq. (\ref{etamat}) one
obtains the $2\times 2$ matrix $\{f^a_P\}$
\begin{equation}
\label{faP}
   \{f^a_P\} =
   \left( \begin{array}{cc}
          f^8_\eta & f^0_\eta \\
          f^8_{\eta'} & f^0_{\eta'}
          \end{array}
   \right) = \left( 
          \begin{array}{cc}
          \cos\theta_p & -\sin\theta_p \\
          \sin\theta_p & \cos\theta_p 
          \end{array}
    \right)
    \left( \begin{array}{cc}
          f^{88} & f^{08} \\
          f^{80} & f^{00} 
          \end{array}
   \right). 
\end{equation}
This matrix depends on four independent parameters $g, M_u, M_s, 
\theta_p$. An alternative parametrization has been considered in
\cite{Kaiser:1997, Leutwyler:1997}, where two constants $f_0,\, f_8$
and two angles $\vartheta_0,\, \vartheta_8$ specify the matrix $f^a_P$. 
\begin{equation}
   \{f^a_P\} =
   \left( \begin{array}{cc}
          f_8\cos\vartheta_8 & -f_0 \sin\vartheta_0 \\
          f_8\sin\vartheta_8 & f_0\cos\vartheta_0
          \end{array}
   \right). 
\end{equation}
 
There is a straightforward correspondence between the parametrization
of Kaiser and Leutwyler and the model predictions. Indeed, one finds 
\begin{equation}
   (f_8)^2=(f_\eta^8)^2 + (f_{\eta'}^8)^2=\frac{1}{3g^2}
   \left(M_u^2+2M_s^2 \right),
\label{f8}
\end{equation}
\begin{equation}
   (f_0)^2=(f_\eta^0)^2 + (f_{\eta'}^0)^2=\frac{1}{3g^2}
   \left(2M_u^2+M_s^2 \right).
\label{f0}
\end{equation}
The formulae for the relations between mixing angles are given in 
the end of this section. 

As one would expect, the model predictions agree well with the general
requirements of chiral symmetry following from chiral perturbation 
theory (ChPT), although the results differ already at lowest order. For 
instance, we have
\begin{equation}
   (f_8)^2=\frac{4f_K^2-f_\pi^2}{3} + \frac{(M_s-M_u)^2}{3g^2}.
\end{equation}
One can show that the second term on the r.h.s. is of order $(m_s-m_u)^2$ 
and therefore must be omitted at lowest order of ChPT. The rest 
of this formula is a well known low-energy relation which is valid in 
standard ChPT. 

The mixing angles $\vartheta_8, \vartheta_0$ are small and 
$\vartheta_8\neq \vartheta_0$. We have for their difference
\begin{eqnarray}
   f_8f_0\sin\left(\vartheta_8 - \vartheta_0\right) 
   &\!\!\! =\!\!\!& f_\eta^8 f_\eta^0 + f_{\eta'}^8 f_{\eta'}^0
   = -\frac{\sqrt{2}}{3g^2}\left(M_s^2-M_u^2\right) \nonumber \\
   &\!\!\! =\!\!\!& -\frac{2\sqrt{2}}{3}\left(f_K^2-f_\pi^2\right)
   -\frac{\sqrt{2}}{6g^2}\left(M_s-M_u\right)^2.
\end{eqnarray}
Thus, $\vartheta_8\neq \vartheta_0$ due to the $SU(3)$ flavour symmetry 
breaking effect. Again, this result agrees with the ChPT formula, if 
one notes that the last term is a higher order contribution.

The singlet decay constant $f_0$ is equal to
\begin{equation}
\label{f0sr}
   (f_0)^2=\frac{2f_K^2+f_\pi^2}{3} 
   + \frac{f_\pi^2}{6}\left(\frac{M_s}{M_u}-1\right)^2.
\end{equation}
In the $\eta'$-extended version of ChPT there is the OZI-rule violating
term in the effective Lagrangian, which contributes as 
$f_\pi^2\Lambda_1$ to the r.h.s. of Eq. (\ref{f0sr}). We have   
instead the term $(M_s/M_u-1)^2/6 \sim \Lambda_1$. Of course, in our
case the origin of this contribution is related with the $SU(3)$
flavour symmetry breaking.

Let us now express the mixing angles $\vartheta_8, \vartheta_0$ in
terms of one mixing angle $\theta_p$ and quark masses. One has
\begin{eqnarray} 
   && \tan\vartheta_8 = \frac{f^8_{\eta'}}{f^8_\eta} = 
      \tan\left(\theta_p - \arctan\frac{\sqrt{2}(M_s-M_u)}{M_u+2M_s}
      \right), \\
   && \tan\vartheta_0 = -\frac{f^0_{\eta}}{f^0_{\eta'}} = 
      \tan\left(\theta_p + \arctan\frac{\sqrt{2}(M_s-M_u)}{2M_u+M_s}
      \right).
\end{eqnarray}
It follows then that 
\begin{eqnarray}
\label{angle8}
   && \vartheta_8 = \psi - \arctan\left(\sqrt{2}\,\frac{M_s}{M_u}
      \right), \\
   && \vartheta_0 = \psi - \arctan\left(\sqrt{2}\,\frac{M_u}{M_s}
      \right),
\label{angle0}
\end{eqnarray}
where $\psi =\theta_p + \arctan\sqrt{2}$. Similar formulae have
been obtained in \cite{FKS:1998} by Feldmann, Kroll, and Stech. 
One should not confuse the angle $\theta_p$, defined by the rotation
(\ref{singoct}) and contributing to the matrix ${f^a_P}$ as it is
shown in Eq. (\ref{faP}), with the naive identification 
$\vartheta_8=\vartheta_0=\theta_p$ discussed in the literature in
connection with the one mixing angle problem. Our consideration is 
perfectly consistent with the two mixing angles approach.

%%%%%%%%%%%%%%%%%%%%%%%%%%%%%%%%%%%%%%%%%%%%%%%%%%%%%%%%%%%%%%%%%%%%%%%%%%
\subsection*{\it 3.3 Strange -- nonstrange basis for decay couplings}
%%%%%%%%%%%%%%%%%%%%%%%%%%%%%%%%%%%%%%%%%%%%%%%%%%%%%%%%%%%%%%%%%%%%%%%%%%
The axial-vector currents can be taken in a different basis, namely,
we shall consider now the nonstrange ${\cal A}_\mu^{ns}$ and strange 
${\cal A}_\mu^{s}$ currents 
\begin{equation}
   {\cal A}_\mu^{ns}=\sqrt{\frac{2}{3}}{\cal A}_\mu^{0}
                    + \frac{1}{\sqrt{3}} {\cal A}_\mu^{8}, \quad
   {\cal A}_\mu^{s}= \frac{1}{\sqrt{3}} {\cal A}_\mu^{0}
                   -\sqrt{\frac{2}{3}}{\cal A}_\mu^{8}.
\end{equation}  

The singlet $\phi^0_R$ and octet $\phi^8_R$ fields are also rotated to
the new basis 
\begin{equation}
\label{qflbas}
   { \phi^{ns} \choose \phi^s } = 
   \frac{1}{\sqrt{3}} \left( 
   \begin{array}{cc}
   \sqrt{2} & 1\\
   1 & -\sqrt{2} 
   \end{array} \right)
   { \phi^0_R \choose \phi^8_R }\, . 
\end{equation}
The corresponding matrix elements are easily calculated 
\begin{equation}
   \langle 0|{\cal A}_\mu^{ns}(0)|\phi^{ns}(p)\rangle 
   =ip_\mu\frac{M_u}{g}\, , \quad
   \langle 0|{\cal A}_\mu^{s}(0)|\phi^{s}(p)\rangle 
   =ip_\mu\frac{M_s}{g}\, .
\end{equation} 

The physical states $P=\eta, \eta'$ are the mixtures of the nonstrange
and strange components, this follows from Eqs. (\ref{singoct}) and 
(\ref{qflbas})
\begin{equation}
\label{str-nonstr}
   { \phi^{ns} \choose \phi^{s} } = \left( 
   \begin{array}{cc}
   \cos\psi & \sin\psi \\
   -\sin\psi & \cos\psi 
   \end{array} \right)
   { \eta \choose \eta' }\, ,
\end{equation}
where the angle $\psi = \theta_p + \arctan \sqrt{2} \simeq \theta_p 
+ 54.74^\circ$.     

Next one can find the couplings describing decays of physical states
in the hadron vacuum
\begin{equation}
   \langle 0|{\cal A}_\mu^{i}(0)|P(p)\rangle =if_P^i p_\mu, \qquad
   (i=ns,s).
\end{equation} 
The result can be represented in a way which is similar to the one of 
Leutwyler -- Kaiser \cite{Feldmann:1999}
\begin{equation}
   \{f^i_P\} =
   \left( \begin{array}{cc}
          f^{ns}_\eta & f^s_\eta \\
          f^{ns}_{\eta'} & f^s_{\eta'}
          \end{array}
   \right) = \left( 
          \begin{array}{cc}
          f_{ns}\cos\vartheta_{ns} & -f_s\sin\vartheta_s \\
          f_{ns}\sin\vartheta_{ns} & f_s\cos\vartheta_s 
          \end{array}
    \right)
\end{equation}
Our calculations show that 
\begin{eqnarray}  
   && f^{ns}_{\eta}=\frac{M_u}{g}\cos\psi , \qquad
      f^{s}_{\eta}=-\frac{M_s}{g}\sin\psi , \nonumber \\
   && f^{ns}_{\eta'}=\frac{M_u}{g}\sin\psi , \qquad
      f^{s}_{\eta'}=\frac{M_s}{g}\cos\psi . 
\end{eqnarray}
It follows that the basic parameters $f_{ns}, f_s, \vartheta_{ns}, 
\vartheta_s$ of the matrix $\{f^i_P\}$, being expressed in terms of
model parameters (in the approximation considered), are
\begin{equation}
   f_{ns}= \frac{M_u}{g}=f_\pi , \quad f_s=\frac{M_s}{g}\, , \quad 
   \psi =\vartheta_{ns}=\vartheta_s.
\end{equation}
There is a direct relation between a common mixing angle
$\vartheta_{ns}=\vartheta_s$ and the OZI-rule which has been discussed 
in \cite{FKS:1998}-\cite{Feldmann:1998}.
 
%%%%%%%%%%%%%%%%%%%%%%%%%%%%%%%%%%%%%%%%%%%%%%%%%%%%%%%%%%%%%%%%%%%%%%%%%%
\subsection*{\it 3.4 Mass formulae and the mixing angle $\theta_p$}
%%%%%%%%%%%%%%%%%%%%%%%%%%%%%%%%%%%%%%%%%%%%%%%%%%%%%%%%%%%%%%%%%%%%%%%%%%
The gap equations are an essential ingredient to obtain mass formulae
of pseudoscalars. One obtains them equating to zero the tadpole 
contributions from the Lagrangian ${\cal L}_{bos}$, see Eq. 
(\ref{Lbos})
\begin{equation}
\label{gap}
  \left\{
       \begin{array}{rcr}
  && h_u+\displaystyle\frac{N_c}{6\pi^2} M_u
         \left(3I_0-\Delta_{us} I_1 \right)=0, \\
  && \\
  && h_s+\displaystyle\frac{N_c}{6\pi^2} M_s
         \left(3I_0+2\Delta_{us} I_1 \right)=0.
        \end{array}
  \right.
\end{equation}

The stationary phase equations (\ref{SPA}) taken in the isospin 
limit ($m_u=m_d$) have been also used to obtain the mass 
spectrum\footnote{See Eq. (\ref{rhoomtau}) for our notations of 
$\omega_i, \rho$ and $\tau_{ij}$.}
\begin{eqnarray}
\label{mpi}
   m_\pi^2 &\!\!\! =\!\!\!& 
           \left(\frac{g^2 m_u}{GM_u}\right)
           \frac{1}{1+\omega_s+\rho+\tau_{uu}}\ , \\
   m_K^2 &\!\!\! =\!\!\!& 
          \frac{g^2}{G}
          \left( \frac{m_u + m_s}{M_u+M_s} \right)
          \frac{1}{1+\omega_u+\rho +\tau_{uu}+\tau_{ss}
          -\tau_{us}}\ , \label{mk} \\
   m_{\eta_\mp}^2 &\!\!\! =\!\!\!&\frac{g^2}{2}\left(A+B\mp
   \sqrt{(A-B)^2+4D^2}\right). \label{masseta} 
\end{eqnarray}
Here $\eta_- =\eta,\, \eta_+ =\eta'$. 

There is a mixing between the $(0,8)$ states in the multiplet. The 
symmetric mass matrix ${\cal M}_p$ is given by
\begin{equation}
\label{massmat}
   {\cal M}_p =  \frac{g^2}{2}\, (\phi^0_R, \phi^8_R) \left( 
                 \begin{array}{cc}
                 A  & D \\
                 D  & B
                 \end{array}
                 \right)
                 {\phi^0_R \choose \phi^8_R}\, ,
\end{equation}
where we have
\begin{eqnarray}
   && A+B=\frac{h_u}{M_u} + \frac{h_s}{M_s} + 
      \frac{2(1+\rho )-\omega_s+\tau_{uu}
      +\tau_{ss}}{G\det N^{(2)}}\, , \\
   && A-B=\frac{1}{3}\left(\frac{h_u}{M_u} - \frac{h_s}{M_s} 
      + \frac{8\omega_u+\omega_s+\tau_{ss}
      -\tau_{uu}}{G\det N^{(2)}}\right)\, , \\
   && D = \frac{\sqrt{2}}{3}\left(
      \frac{h_u}{M_u} - \frac{h_s}{M_s} 
      + \frac{\omega_s-\omega_u +\tau_{ss}
      -\tau_{uu}}{G\det N^{(2)}}\right).
\end{eqnarray}
The non-diagonal term of the matrix vanishes in the $SU(3)$ flavour 
symmetric case $(m_u=m_d=m_s)$, otherwise $D\neq 0$. This matrix is 
diagonalized by an orthogonal transformation (\ref{singoct}) to the 
states $(\eta , \eta' )$ with masses given by Eqs. (\ref{masseta}).
The singlet-octet mixing angle is 
\begin{equation}
\label{anglep}
   \tan 2\theta_p =\frac{2D}{A-B}\, .
\end{equation}

It is easily seen from Eq. (\ref{anglep}) that the mixing angle
$\theta_p$ is equal to its ideal value: $\tan (2\theta_\id
)=2\sqrt{2}$, i.e., $\theta_\id\simeq 35.26^\circ$ and $\psi = 
90^\circ$, if $\kappa =0$. As a result we have\footnote{Note that 
the orthogonal transformation (\ref{singoct}) is written for  
$\kappa\neq 0$, what corresponds to $A-B>0$. If $\kappa =0$, 
one has the opposite inequality $A-B<0$, and, as a consequence, one 
should replace in Eqs. (\ref{singoct}) and (\ref{str-nonstr}) the 
fields $\eta\leftrightarrow\eta'$.} $\eta\propto \phi^{ns}$ and 
$\eta'\propto\phi^s$, thus the eight-quark interactions have no 
inf\mbox{}luence on the f\mbox{}lavour content of $\eta, \eta'$
without the 't Hooft term. 

Consider now the $\eta -\eta'$ masses (\ref{masseta}) presented as 
follows
\begin{equation}
   m_{\eta_\mp}^2 = m_K^2+Q_1  
   \mp\sqrt{\left(m_K^2-m_\pi^2-Q_2\right)^2 +2Q_3^2}\, . 
\label{etama} 
\end{equation}
The independent functions $Q_1,\, Q_2,\, Q_3$ are equal to
\begin{eqnarray} 
\label{Q1}
   Q_1 &\!\!\! =\!\!\! & \frac{g^2(2\omega_u +\omega_s +\tau_{uu}
               + \tau_{ss} -2\tau_{us})}{2G \det N^{(2)}}
               -\left(M_s-M_u\right)^2  \nonumber \\
       &\!\!\! -\!\!\! & \frac{2g^2(\omega_u -\tau_{us})
       (\omega_s -\omega_u +\tau_{uu}+\tau_{ss}-2\tau_{us})
       }{G(1+\rho +\omega_u +\tau_{uu}+\tau_{ss}-\tau_{us}) 
       \det N^{(2)}}\, , \\ 
   Q_2 &\!\!\! =\!\!\! & \frac{g^2(\omega_s +\tau_{ss}
               - \tau_{uu})}{2G \det N^{(2)}}
               +\left(M_s-M_u\right)^2  \nonumber \\
       &\!\!\! +\!\!\! & \frac{g^2 (\omega_s -\omega_u 
       +\tau_{us}- \tau_{ss})}{G(1+\rho +\omega_s +\tau_{uu}) 
       (1+\rho +\omega_u +\tau_{uu}+\tau_{ss}-\tau_{us} )}\, , 
       \label{Q2} \\
   Q_3 &\!\!\! =\!\!\! &\frac{g^2\omega_u}{G\det N^{(2)}}\, . 
\label{Q3}
\end{eqnarray}
In the large-$N_c$ approximation ($Q_1, Q_2, Q_3\to Q_1^{LO},
Q_2^{LO}, Q_3^{LO}$) these functions are related to each 
other and to the ghost coupling $\lambda_\eta$, introduced by
Veneziano in \cite{Witten:1979}, 
\begin{equation}
   \frac{2}{3}Q_1^{LO} = 2Q_2^{LO} = Q_3^{LO} = \frac{g^2\omega}{G} 
   =\frac{\lambda_\eta^2}{N_c} 
   \qquad (\mbox{large}\ N_c),
\end{equation}
where $\omega \sim 1/N_c$ is a leading order contribution of
$\omega_u$, or $\omega_s$. We have for $\lambda_\eta^2$
\begin{equation}
\label{lncl}
   \lambda^2_\eta =-\frac{\kappa N_c}{16f_\pi^2}
   \left.\left(\frac{M}{G}\right)^3\right|_{N_c\to\infty}.
\end{equation}
With these specific values of $Q$'s our expressions for the  masses of 
$\eta,\, \eta'$ mesons coincide with Eq. (34) of Veneziano work in 
\cite{Witten:1979}. Moreover, the Witten -- Veneziano formula for 
the mass of $\eta'$, 
\begin{equation}
\label{WV1}
   m_{\eta'}^2+m_\eta^2-2m_K^2 
   =\lambda^2_\eta = -\frac{6}{f_\pi^2}\chi (0)|_{YM},
\end{equation}
which is obtained in the large $N_c$ limit of QCD for non-vanishing 
quark masses, relates the $\eta'$ mass with the topological 
susceptibility in pure Yang -- Mills theory $\chi (0)|_{YM}$. 
Consequently, Eq. (\ref{lncl}) yields 
\begin{equation}
\label{tops}
   \chi (0)|_{YM} = \frac{\kappa N_c}{12}
   \left(\frac{M}{2G}\right)^3
   \qquad (\mbox{large}\ N_c).
%\right|_{N_c\to\infty}.
\end{equation} 
The interesting feature here is the explicit demonstration that the 
eight-quark forces contribute to Eqs. (\ref{Q1})-(\ref{Q3}) in such a 
way that the only dominant term in (\ref{WV1}) is still the 't Hooft 
interaction, even though the $\rho_{ij},\, \tau_{ij}$ may formally 
be of the same $1/N_c$-order as $\omega_i$.   

One can use the most recent lattice calculation of the topological 
susceptibility in \cite{Del:2005}: $\,\chi (0)|_{YM} = - (1.33 \pm
0.14) \times 10^{-3}\, \mbox{GeV}^4$ to find $Q_2^{LO}=0.156\, 
\mbox{GeV}^2$. Additionally, we have at leading $N_c$-order 
\begin{equation}
\label{anglepes}
   \tan 2\theta_p = 2\sqrt{2}\, \frac{m_K^2- m_\pi^2 - Q_2 
   +\frac{1}{2}Q_3}{m_K^2-m_\pi^2 -Q_2-4Q_3} \to
   2\sqrt{2}\, \frac{m_K^2- m_\pi^2}{m_K^2-m_\pi^2 -9Q_2^{LO}}, 
\end{equation}
obtaining approximately a mixing angle $\theta_p\simeq -14^\circ$, and
masses $m_\eta\simeq 516\,\mbox{MeV}$, $m_{\eta'}\simeq
1077\,\mbox{MeV}$. These numbers reflect the general picture
presented in Section 5 rather well.

%%%%%%%%%%%%%%%%%%%%%%%%%%%%%%%%%%%%%%%%%%%%%%%%%%%%%%%%%%%%%%%%%%%%%%%%%%%%%%

\section*{\centerline{\large\bf 4. Scalars: masses and the 
          mixing angle $\theta_s$}}

%%%%%%%%%%%%%%%%%%%%%%%%%%%%%%%%%%%%%%%%%%%%%%%%%%%%%%%%%%%%%%%%%%%%%%%%%%%%%%

%%%%%%%%%%%%%%%%%%%%%%%%%%%%%%%%%%%%%%%%%%%%%%%%%%%%%%%%%%%%%%%%%%%%%%%%%%
\subsection*{\it 4.1 Mass spectrum of the scalar nonet}
%%%%%%%%%%%%%%%%%%%%%%%%%%%%%%%%%%%%%%%%%%%%%%%%%%%%%%%%%%%%%%%%%%%%%%%%%%
The masses of the scalar nonet:  $a_0\ (I=1)$, $K^*_0\ (I=1/2)$,  
$f_0^\mp\ (I=0),$ are  
\begin{eqnarray}
   && m_{a_0}^2 = m_\pi^2 + 4M_u^2 
   + \frac{2g^2(\omega_s - \tau_{uu})}{G[
   (1+\rho + 2\tau_{uu})^2 - (\omega_s
   -\tau_{uu})^2]}\, , \\
   && m_{K_0^*}^2 = m_K^2 + 4 M_u M_s 
   + \frac{ 2g^2(\omega_u- \tau_{us})
   }{G[(1+\rho + \tau_{uu} + \tau_{ss})^2 
   - (\omega_u - \tau_{us})^2]}
   \label{K0*}\, , \\
   && m^2_{f_0^\mp} = \frac{g^2}{2} \left( {\cal A} 
   + {\cal B} \mp \sqrt{({\cal A} - {\cal B} )^2
   + 4 {\cal D}^2}\right). 
\label{f0mp}
\end{eqnarray}

To obtain the last formula (\ref{f0mp}) the mass matrix of isospin 
singlets $\sigma^0_R$ and $\sigma^8_R$ in the Lagrangian (\ref{mass})
\begin{equation}
\label{scmasses}
   {\cal M}_s =  \frac{g^2}{2}\, (\sigma^0_R, \sigma^8_R) 
                 \left( \begin{array}{cc}
                 {\cal A}  & {\cal D} \\
                 {\cal D}  & {\cal B}
                 \end{array} \right)
    {\sigma^0_R \choose \sigma^8_R}\, ,
\end{equation}
where
\begin{eqnarray}
   && {\cal A} + {\cal B} =
   \frac{h_u}{M_u} + \frac{h_s}{M_s} + 
   \frac{N_c I_1}{\pi^2}\,\left( M_s^2+M_u^2 \right)
   + \frac{2+\omega_s +4\rho +3 ( \tau_{uu} +\tau_{ss}) }{
   G\det N^{(1)}}\, , \nonumber\\
   && {\cal A} - {\cal B} =
   \frac{h_u}{M_u} - \frac{h_s}{M_s} - 
   \frac{8\omega_u+\omega_s+ 2 (2\rho_{uu} - \rho_{ss} +8\rho_{us})
   + 3 ( \tau_{uu} - \tau_{ss}) }{
   3G\det N^{(1)}}\, , \nonumber\\
   && {\cal D} = \sqrt{2} \left(
   \frac{h_u}{M_u} -\frac{h_s}{M_s}
   -\frac{\omega_s-\omega_u + 2 (2\rho_{uu} - \rho_{ss} - \rho_{us}) 
   + 3 ( \tau_{uu} - \tau_{ss}) }{3G\det N^{(1)}}\right), \nonumber
\end{eqnarray}
has been diagonalized by an orthogonal transformation 
\begin{equation}
\label{ortog}
   {f_0^- \choose f_0^+ } =\left( 
   \begin{array}{cc}
   \cos\theta_s  & \sin\theta_s \\
   -\sin\theta_s & \cos\theta_s
   \end{array} \right)
   {\sigma^0_R \choose \sigma^8_R}\, ,
\end{equation}
with the angle given by
\begin{equation}
\label{anglsc}
   \tan 2\theta_s =\frac{2{\cal D}}{{\cal A} -{\cal B}}\, .
\end{equation}

Let us apply formula (\ref{anglsc}) to the case in which the 't Hooft 
determinant is neglected. We f\mbox{}ind that the mixing angle
$\theta_s\neq\theta_\id$ at $\kappa =0$. The reason is that the terms 
which are proportional to $\rho_{us}$ in ${\cal D}$ and 
${\cal A}-{\cal B}$ have different coefficients. Therefore the $f_0^-$ 
meson has an admixture of the strange component and correspondingly
the $f_0^+$ meson has an admixture of nonstrange quarks due to the 
eight-quark interactions\footnote{We have $f_0^-\propto (\bar
uu+\bar dd)$ and $f_0^+\propto \bar ss$ at $\kappa = g_1 =0$.}
with coupling $g_1$. Such admixtures explicitly violate the OZI rule
in these scalar channels. 

We must notice that the singlet-octet splitting of scalars is more 
sensitive to the eight-quark interactions, as opposed to the 
pseudoscalar case considered above. We can gain some understanding of 
this by writing slightly less explicit formulae for scalars. For
that we turn again to the large $N_c$ arguments, postponing the exact 
calculations till Section 5. 
 
%%%%%%%%%%%%%%%%%%%%%%%%%%%%%%%%%%%%%%%%%%%%%%%%%%%%%%%%%%%%%%%%%%%%%%%%%%
\subsection*{\it 4.2 The $1/N_c$ consideration}
%%%%%%%%%%%%%%%%%%%%%%%%%%%%%%%%%%%%%%%%%%%%%%%%%%%%%%%%%%%%%%%%%%%%%%%%%%
The $1/N_c$ expansion is usually a good approximation for hadrons. 
If we accept this idea, we can deduce from the above formulae a clear
qualitative picture of the role played by the eight-quark forces in 
the mass spectra of scalars. Our starting point are the following 
expressions
\begin{eqnarray}
   m_{a_0}^2 &\!\!\! =\!\!& m_\pi^2 + 4M_u^2 
   + \frac{2g^2}{G}\left(\omega_s - \tau_{uu}\right)
   + \ldots\, , 
   \label{a0exp} \\
   m_{K_0^*}^2 &\!\!\! =\!\!& m_K^2 + 4 M_u M_s 
   + \frac{2g^2}{G}\left(\omega_u - \tau_{us}\right)
   + \ldots\, , 
   \label{K0*exp} \\
   m^2_{f_0^\mp} &\!\!\! =\!\!& m_K^2+(M_s+M_u)^2+{\cal Q}_1 
   \nonumber \\ 
   &\!\!\!\mp\!\!& \sqrt{(\Delta_{K\pi} + {\cal Q}_2 )^2 
   + 8 (\Delta_{K\pi} + {\cal Q}_3)^2}\, , 
\label{f0mp2}
\end{eqnarray}
where $\Delta_{K\pi}=m_K^2-m_\pi^2$, and ellipses are used 
to denote omitted terms of $1/N_c^2$ order and higher. The functions 
${\cal Q}_1, {\cal Q}_2, {\cal Q}_3$ are
\begin{eqnarray}
\label{R1}
   && {\cal Q}_1 = \frac{g^2}{2G}\left[2\omega_u-\omega_s -2\rho 
          - (\tau_{uu}+\tau_{ss}+2\tau_{us})\right]
          + \ldots\, , \\
   && {\cal Q}_2 = -Q_2^{LO}+\frac{g^2}{3G}\left( 4\omega_u +2\omega_s 
          + 2\rho_{uu} -\rho_{ss} + 8\rho_{us} \right) 
          + \ldots\, , 
\label{R2} \\
   && {\cal Q}_3 =-Q_2^{LO}
          + \frac{g^2}{3G}\left( 2\omega_s -\frac{\omega_u}{2} 
          + 2\rho_{uu} -\rho_{ss} -\rho_{us} \right) 
          + \ldots\, 
\label{R3}.  
\end{eqnarray}
The contributions from the different $\tau$'s are exactly canceled 
in Eqs. (\ref{R2})-(\ref{R3}) at this order. Moreover, note that the 
$SU(3)$ breaking corrections have a higher order in $1/N_c$. For 
instance, $\omega_u, \omega_s\sim 1/N_c$, but the difference $\omega_u
-\omega_s\sim 1/N_c^2$. The same is true for other flavour dependent 
functions in Eqs. (\ref{R1})-(\ref{R3}). Thus, one has to lowest order 
in $1/N_c$ 
\begin{equation} 
   {\cal Q}_1^{LO}=Q_2^{LO}-3E_1^{LO}-2E_2^{LO}, \quad 
   {\cal Q}_2^{LO}=3\left(Q_2^{LO}+E_1^{LO}\right), \quad 
   {\cal Q}_3^{LO}= 0,
\end{equation}   
where the eight-quark contributions $E_1^{LO}$ and $E_2^{LO}$, namely 
\begin{equation}
   E_1^{LO} = \left.\frac{g^2\rho_{uu}}{G}\right|_{large\, N_c}, 
   \qquad
   E_2^{LO} = \left.\frac{g^2\tau_{uu}}{G}\right|_{large\, N_c},
\end{equation} 
are proportional to the strengths $\sim g_1$ and $g_2$ correspondingly. 

Next, let us try to understand in simple terms the hierarchy inside 
the nonet. It is easy to see from Eqs. (\ref{a0exp})-(\ref{f0mp2}) that
$m_{f_0^-}<m_{a_0}<m_{K_0^*}<m_{f_0^+}$. This is in an agreement with
the result of analysis \cite{Dmitr:1996}. Let us do some crude 
numerical estimates. We have at leading order\footnote{The last
Eq. (\ref{K-pi}) follows from our general result 
Eqs. (\ref{mpi})-(\ref{mk}) in the pseudoscalar sector for the difference 
$$
   m_K^2-m_\pi^2 =2M_s(M_s-M_u)+\frac{g^2(\omega_s-\omega_u+\tau_{us}
   -\tau_{ss})}{G(1+\omega_s+\rho + \tau_{uu})(1 + \omega_u + \rho 
   +\tau_{uu} +\tau_{ss}-\tau_{us})}\, .
$$}
\begin{eqnarray}
   && m^2_{K^*_0}-m^2_{f_0^\mp} = {\cal Q}_2^{LO} \pm
      \sqrt{(\Delta_{K\pi}+{\cal Q}_2^{LO})^2
      +8\Delta^2_{K\pi}}, \label{s-of0}\\
   && m^2_{K^*_0}-m^2_{a_0}=\Delta_{K\pi}+4M_u(M_s-M_u), 
   \label{K0*-a0} \\
   && \Delta_{K\pi}=2M_s(M_s-M_u). \label{K-pi}
\end{eqnarray}

First, we conclude that eight-quark forces are probably unimportant 
for the difference $m^2_{K^*_0}-m^2_{a_0}$. 

Second, we use the ratio $f_K/f_\pi =1.22$ to find $M_s/M_u$. Indeed, 
it follows from Eqs. (\ref{GT}), (\ref{f8}), (\ref{f0}), (\ref{angle8}) 
and (\ref{angle0}) that    
\begin{eqnarray}  
   && \frac{f_K}{f_\pi}=\frac{1}{2}\left(1+\frac{M_s}{M_u}\right), \\
   && \frac{f_8}{f_\pi} = \sqrt{\frac{1}{3} \left(1 + 
      2\frac{M_s^2}{M_u^2}\right)}\, , \qquad 
      \frac{f_0}{f_\pi} = \sqrt{\frac{1}{3} \left(2 + 
      \frac{M_s^2}{M_u^2}\right)}\, , \\ 
   && \vartheta_0 -\vartheta_8 = \arctan\left[\frac{\sqrt{2}}{3}
      \left(\frac{M_s}{M_u}-\frac{M_u}{M_s}\right)\right]. 
\end{eqnarray}
One easily finds   $M_s/M_u=1.44$. Numerically, this yields $f_8=1.31 
f_\pi$, $f_0= 1.17 f_\pi$, $\vartheta_0 -\vartheta_8 = 19.5^\circ$.

Third, we use phenomenological values $m_\pi \simeq 138$ MeV 
(averaged over the isotopic triplet $\pi^0,\, \pi^\pm$) and 
$m_K\simeq 495.7$ MeV (averaged value for isotopic duplet $K^+,\, 
K^0$) to obtain $\Delta_{K\pi}=0.227\,\mbox{GeV}^2$, and
$m^2_{K^*_0}-m^2_{a_0}\simeq 2.39 \Delta_{K\pi}=0.542\,\mbox{GeV}^2$.

In order to make more progress, we need a further dynamical input. For 
that we identify the quark-antiquark $a_0$-state of the model with the 
known $I^G(J^{PC})=1^-(0^{++})$ resonance $a_0(980)$. This resonance is 
often considered as $K\bar K$ molecular-like bound state 
\cite{Isgur:1990}. The four-quark nature of the $a_0(980)$ meson is 
also widely discussed in the literature (see, e.g., the recent paper 
\cite{Achasov:2003} and references therein), where $a_0(980)$ is a
compact $K\bar K$ state. The extended molecule case does not exclude 
that the core part of the wave function may be dominantly $q\bar q$ 
\cite{Klempt:2004}. The four-quark picture is  however based essentially 
on the MIT-bag model, thus representing an alternative 
approach to the problem.
  
Now, by using that $m_{a_0(980)}\simeq 980\, \mbox{MeV}$, one derives 
$m_{K_0^*}\simeq 1226\,\mbox{MeV}$. We suppose that this state may be 
identified with the wide $I(J^{P})=\frac{1}{2}(0^{+})$ resonance 
$K_0^*(800)$: $m_{K_0^*(800)} = 797\pm 19\pm 43\ \mbox{MeV}$,
$\,\Gamma = 410\pm 43\pm 87\ \mbox{MeV}$, reported in 
\cite{Aitala:2002}; see also \cite{Goebel:2000}-\cite{Pelaez:1998}, 
where further support for this low lying state is given.

Before we calculate the masses of the two $f_0^\mp$ states from 
Eq. (\ref{s-of0}), consider the singlet-octet mixing angle $\theta_s$ 
which can be written very compactly in the large $N_c$ world, viz., 
\begin{equation} 
\label{anglescep}
   \tan 2\theta_s \simeq 2\sqrt{2}\, 
   \frac{m_K^2- m_\pi^2}{m_K^2-m_\pi^2 +{\cal Q}_2^{LO}}\, . 
\end{equation}
If the 't Hooft interaction term $Q_2^{LO}$ dominates over $E_1^{LO}$ 
in ${\cal Q}_2^{LO}$ (this is possible when $g_1\sim 1/N_c^5$; 
in this case $E_1^{LO}\sim 1/N_c^2 \ll Q_2^{LO}\sim 1/N_c$), one
easily finds that $\theta_s\simeq 21^\circ$ (we have used here the 
estimate $Q_2^{LO}=0.156\,\mbox{GeV}^2$ obtained before 
Eq. (\ref{anglepes})). 

One immediately derives from Eq. (\ref{s-of0}) the masses of the 
singlet-octet mixed states $f_0^{\mp}$: $\, m_{f_0^-}\simeq 300\, 
\mbox{MeV}$ and $m_{f_0^+}\simeq 1407\, \mbox{MeV}$. 

There is a strong cancellation in the formula for the $f_0^-$-mass
\begin{eqnarray}
\label{f0approx}
   m_{f_0^-}^2&\!\!\!\simeq\!\!\!& m_{K_0^*}^2-3Q_2^{LO}
   -\sqrt{(\Delta_{K\pi}+3Q_2^{LO})^2+8\Delta_{K\pi}^2} 
   \nonumber \\
   &\!\!\!\simeq\!\!\!&
   (1.502-0.468-0.946=0.088)\, \mbox{GeV}^2. 
\end{eqnarray} 
As a consequence the result is very sensitive to the parameters of the model. 
We shall see later that the best fit of the pseudoscalar channel leads
to the value $m_{f_0^-}\simeq 550-750\, \mbox{MeV}$. Therefore, the 
lowest mass scalar meson, $f_0^-$, is identified with $f_0(600)$. The
Particle Data Group assigns to this resonance the mass $m_{f_0(600)} 
= 400-1200\ \mbox{MeV}$, and the width $\Gamma = 600-1000\ \mbox{MeV}$. 

The state $f_0^+$ agrees with the state $f_0(1370)$: $m_{f_0(1370)}
=1200-1500\,\mbox{MeV}$ ($\Gamma =200-500\, \mbox{MeV}$). The lower 
state $f_0(980)$ with the same quantum numbers is too far away from 
the value following from Eq. (\ref{s-of0}), thus our estimate shows 
that $f_0(980)$ may be not a member of the scalar quark-antiquark 
nonet considered\footnote{It is notorious that the low-lying scalars 
are still the subject of many studies. Our conclusion agrees with some 
other results \cite{Isgur:1990,Achasov:2003,Uehara:2004}, but we are 
aware that the point requires an additional analysis of scalar decays
to be definitive. In different approaches, based on a coupled channel analysis, 
the pole position may also be affected by closed channels, see e.g. \cite{Beveren:}.}

%%%%%%%%%%%%%%%%%%%%%%%%%%%%%%%%%%%%%%%%%%%%%%%%%%%%%%%%%%%%%%%%%%%%%%%%%%
%\subsection*{\it 4.3 Sum rule for pseudoscalar and scalar nonets}
%%%%%%%%%%%%%%%%%%%%%%%%%%%%%%%%%%%%%%%%%%%%%%%%%%%%%%%%%%%%%%%%%%%%%%%%%%
It is known that the $U(1)_A$ breaking produces an opposite in sign 
mass-squared splitting between the octet and the singlet for the 
scalar and pseudoscalar mesons. This can be embodied in the 
approximate sum rule \cite{Dmitr:1996,Petry:1995}
\begin{equation}
   m_{\eta'}^2+m_\eta^2-2m_K^2 + m_{f_0^+}^2 + 
   m_{f_0^-}^2-2m_{K_0^*}^2 \simeq 0. 
\end{equation}
The r.h.s. results from the exact cancellation between the $1/N_c$ 
order terms induced by the 't Hooft interaction (see, for example, 
Eq. (\ref{WV1})). One can see that eight-quark interactions may 
contribute to the sum rule already at $1/N_c$ order, if the coupling $g_1$ 
counts as $1/N_c^4$. Indeed, one obtains    
\begin{equation}
   m_{\eta'}^2+m_\eta^2-2m_K^2 + m_{f_0^+}^2 + 
   m_{f_0^-}^2-2m_{K_0^*}^2 = - 6E_1^{LO}
   +{\cal O}\left(\frac{1}{N^2_c}\right). 
\end{equation}
This term has a negative sign, decreasing the sum $m_{f_0^-}^2 + 
m_{f_0^+}^2$. Let us note that in this case $g_1$ lowers the value of 
$m_{f_0^-}$ and due to the fine tuning effect in Eq. (\ref{f0approx}) 
the octet-singlet splitting grows with increasing $g_1$ in the scalar 
nonet. Thus the above sum rule is a good illustration of the possible 
impact of the eight-quark OZI violating forces on the scalar mesons.

%%%%%%%%%%%%%%%%%%%%%%%%%%%%%%%%%%%%%%%%%%%%%%%%%%%%%%%%%%%%%%%%%%%%%%%%%%%%%%

\section*{\centerline{\large\bf 5. Numerical results}}

%%%%%%%%%%%%%%%%%%%%%%%%%%%%%%%%%%%%%%%%%%%%%%%%%%%%%%%%%%%%%%%%%%%%%%%%%%%%%%

%%%%%%%%%%%%%%%%%%%%%%%%%%%%%%%%%%%%%%%%%%%%%%%%%%%%%%%%%%%%%%%%%%%%%%%%%%
%\subsection*{\it 5.2 Data in the pseudoscalar channel}
%%%%%%%%%%%%%%%%%%%%%%%%%%%%%%%%%%%%%%%%%%%%%%%%%%%%%%%%%%%%%%%%%%%%%%%%%%
We collect the results of the exact numerical calculations for the 
mass-spectra, mixing angles and quark condensates in three tables. 
Input is denoted by a *. Table 1 contains the seven parameters of 
the model $m_u,\, m_s,\, \Lambda,\, G,\, \kappa,\, g_1,\, g_2$. 

Sets $({\rm a,b,c})$ and $({\rm d,e,f})$, are each a block for which 
we compare the effect of the new parameters $g_1,\, g_2$ as follows. 
In the first line of each grouping we set $g_1,\, g_2$ to zero, and 
fit four of the remaining parameters ($m_u, m_s, G, \kappa$) by 
fixing $m_\pi,\, m_K, \, f_\pi,\, f_K$; in the first set $\Lambda$ is 
fixed through $f_0^-$, in the second through $\eta'$. The reason to
fix the empirically not well known mass of $f_0^-$ is that it is the 
most sensitive to changes of the parameter $g_1$ (we remind our
discussion of Eq. (\ref{f0approx})). By fixing it, one reverts the 
situation and is able to detect the effects of $g_1,\, g_2$ on the 
other observables. In the last set one sees that by fixing the mass 
of $\eta'$, it is $f_0^-$ that monopolizes the value of $g_1$. 

\vspace{0.5cm}
\noindent
{\small Table 1 \\
\noindent Parameters of the model: $m_u,\, m_s$ (MeV), $G$
(GeV$^{-2}$), $\Lambda$ (MeV), $\kappa$ (GeV$^{-5}$), $g_1,\, g_2$ 
(GeV$^{-8}$). We also show the corresponding values of constituent 
quark masses $M_u$ and $M_s$ (MeV).}
\vspace{0.2cm}

\noindent
\begin{tabular}{lcccccccccc}
\hline
&$m_u$  &$m_s$  &$M_u$  &$M_s$  &$\Lambda$ &G   &$-\kappa$    
&$g_1$   &$g_2$   
\\ \hline
a & 5.2 & 161 & 302 & 486 & 934  & 7.18  & 1122  & 0*    &0*      \\
b & 5.5 & 175 & 325 & 523 & 896  & 8.78  &  774  & 1000* &0*      \\ 
c & 5.4 & 173 & 322 & 519 & 900  & 8.57  &  822  & 1000* &-132*   \\ 
d & 6.1 & 189 & 372 & 646 &  839 & 12.16 & 1082  & 0*    &0*      \\ 
e & 6.1 & 189 & 372 & 646 &  839 & 11.28 & 1083  &1500*  &327.24  \\ 
f & 6.1 & 189 & 372 & 646 &  839 & 8.92  & 1083  &6000*  & 327.24 \\ 
\hline
\end{tabular}   
\vspace{0.7cm}

For all sets, except of course the case $g_1=g_2=0$, the stability 
conditions (\ref{ineq1}) are fulfilled.   

Tables 2 and 3 contain the results to be compared with the
phenomenological data. The first set shows drastic effects of the 
parameter $g_1$ on mass spectra: going from $({\rm a})$ to $({\rm b})$ 
the $\eta'$ mass is reduced by $40\%$, getting close to its empirical 
value $m_{\eta'}=957.78\pm 0.14\, \mbox{MeV}$, while the $\eta$ mass 
gets smaller by $7\%$. The overall effect is a reduction of the gap 
between these two states, which can be translated into the smaller 
value of the parameter $\kappa$ in $({\rm b})$, as compared to the one 
in $({\rm a})$. 

The $\eta -\eta'$ splitting can be illustrated by the formula 
(\ref{etama}) which takes the form
\begin{equation}
\label{Venext}
   m_\eta^2 = m_0^2  
   -\frac{8(m_K^2-m_\pi^2)^2+3c_q}{9(m_{\eta'}^2-m_0^2)}\, ,  
\end{equation}
where $m_0^2=\frac{1}{3}(4m_K^2-m_\pi^2)$ is the Gell-Mann 
-- Okubo result for the $\eta$-mass. The remainder originates in the 
repulsion of $\eta$ and $\eta'$ and represents an $SU(3)$ breaking
effect of second order. The coefficient $c_q$ depends on the $Q$'s given 
by Eqs. (\ref{Q1})-(\ref{Q3}), namely, $c_q=2(m_K^2-m_\pi^2)(Q_1-3Q_2) 
+3(Q_2^2-Q_1^2+2Q_3^2)$. Formula (\ref{Venext}) for $c_q\neq 0$
extends the Veneziano result \cite{Witten:1979} (see Eqs. (34)
there) by including the $SU(3)$ breaking corrections stemming from the 
't Hooft and eight-quark interactions.

Numerically $m_0=565\,\mbox{MeV}$ is just a little bit larger, to be 
compared with the phenomenological value $m_{\eta} = 547.30\pm 0.12\, 
\mbox{MeV}$. The Witten-Veneziano correction (the second term $\sim 
(m_K^2-m_\pi^2)^2$) is related to the topological susceptibility\footnote{
One can find details, in particular, in the first reference of 
\cite{Leutwyler:1997}.} and is about four times larger than it is 
required. Considering set ({\rm b}), we obtain $m_\eta \simeq 496\,
\mbox{MeV}$. This value is now corrected by the six and eight-quark 
contributions collected in $c_q$. Unfortunately they work in the same 
direction and do not improve the low result for $\eta$. Finally one 
finds the value $m_\eta =486\,\mbox{MeV}$. This is the general tendency 
in all sets.    

\vspace{0.5cm}
\noindent
{\small Table 2 \\
\noindent The masses, weak decay constants of light pseudoscalars 
(in MeV), the singlet-octet mixing angle $\theta_\p$ (in degrees), 
and the quark condensates $\big <\bar uu\big >, \big <\bar ss\big >$ 
expressed as usual by positive combinations in MeV. 
}
\vspace{0.2cm}

\noindent
\begin{tabular}{lcccccccccc}
\hline
& $m_\pi$ & $m_K$  & $m_\eta$ & $m_{\eta'}$ & $f_\pi$ & $f_K$ 
& $\theta_\p $&$-\big <\bar uu\big >^{\frac{1}{3}}$ 
&$-\big <\bar ss\big >^{\frac{1}{3}}$
 \\ 
\hline
a \hspace{5cm}& 138* & 494* & 525 & 1761 & 92* & 120*   & -1.& 246 & 210  \\ 
b & 138* & 494* & 486 & 968 & 92* & 120*   & -12 & 242 & 199 \\ 
c & 138* & 494* & 493 & 1023 & 92* & 120*   & -10 & 242 & 200  \\ 
d & 138* & 494* & 476 & 958* & 92* & 116*    & -14.4 & 233 &184  \\ 
e & 138* & 494* & 476 & 958* & 92* & 116*    & -14.4 & 233 & 184 \\ 
f & 138* & 494* & 476 & 958* & 92* & 116*    & -14.4 & 233&184  \\ 
\hline
\end{tabular}
\vspace{0.7cm}

Other effects are seen in the reduction of all scalar masses, except 
the input one, by $25\%$ for $a_0$, $14\%$ for $K_0^*$, $9\%$ for  
$f_0^+$. In set $({\rm c})$ the effect of $g_2$, keeping the same 
$g_1$ as in set $({\rm b})$, is seen to increase all masses again. It 
has in this case a negative value, which is also allowed by the 
stability conditions. The condensates and scalar mixing angle remain 
comparable in $({\rm a,b,c})$, the pseudoscalar angle $\theta_p$ 
increases in absolute value from set $({\rm a})$ to $({\rm b})$. 
Its value quoted in Table 2 is around $\theta_p=-12^\circ$ being in
agreement with estimates of Veneziano \cite{Witten:1979} and more 
recent calculations in \cite{Scadron:1990}.  

This angle is directly related by Eqs. (\ref{angle8})-(\ref{angle0}) 
with two mixing angles seen in the singlet and the octet components of 
the decay constants. We find, for example, for set ({\rm c}): 
$\vartheta_8=-21.6^\circ,\, \vartheta_0=3.5^\circ$. 

There is some ambiguity about the definition of quark condensates if 
the chiral symmetry is explicitly broken by bare quark masses $m_i$. 
The values given in Table 2 are obtained by the subtraction of the 
expectation value of $\bar q_iq_i$ in the perturbative vacuum from 
its expectation in the true vacuum:
\begin{equation}
   \big <\bar q_iq_i\big > =\frac{1}{2}\left(
   \left.{\rm h}_i\right|_{\Delta_i\neq 0} - 
   \left.{\rm h}_i\right|_{\Delta_i=0}\right),  
\end{equation}
which is the definition used in \cite{Bernard:1988}. Let us recall the
recent update of the light-quark condensate at a scale of $1$ GeV: 
$\big <\bar qq\big >(1\ \mbox{GeV})=-(242\pm 15\ \mbox{MeV})^3$, where 
$\bar qq=(\bar uu+\bar dd)/2$ represents the isospin average of the 
non-strage quarks \cite{Jamin:2002}. The f\mbox{}lavour breaking ratio 
is known to be $\big <\bar ss\big >/\big <\bar qq \big >=0.8\pm 0.3$ 
\cite{Jamin:2002}. 

The second set $({\rm d,e,f})$: we fix the $\eta'$ mass to its 
empirical value, and chose also $f_K$ closer to experiment, keeping 
$m_\pi, m_K, f_\pi$ as in all other cases. In $({\rm e})$ we obtain 
$g_2$ through the mass of $a_0(980)$ and take $g_1$ arbitrarily. 
In the pseudoscalar sector this lowers slightly the value of the 
$\eta$-mass and changes a bit the mixing angle $\theta_p$. 

\vspace{0.5cm}
\noindent
{\small Table 3 \\
\noindent The masses of the scalar nonet (in MeV), and the 
corresponding singlet-octet mixing angle $\theta_\s$ (in degrees). 
}
\vspace{0.2cm}

\noindent
\begin{tabular}{lccccc}
\hline
& $m_{a_0(980)}$ & $m_{K_0^*(800)}$ & $m_{f_0(600)}$ 
& $m_{f_0(1370)}$ & $\theta_\s $ \\ 
\hline
a & 1262  & 1347  & 600*  & 1436 & 16 \\ 
b & 945  & 1150  & 600*  & 1309 & 15 \\ 
c & 980  & 1176  & 600*  & 1326 & 21 \\ 
d & 993 & 1217 & 754  & 1391 & 25 \\ 
e & 980* & 1204  & 691  & 1374 & 23 \\ 
f & 980* & 1204  & 559  & 1362 & 23  \\ 
\hline
\end{tabular}
\vspace{0.7cm}

The main effect is visible in the $f_0(600)$ mass: a further increase 
in $g_1$, set $({\rm f})$, decreases further its mass, leaving the 
remining observables almost unaffected. The repulsion between the 
two isosinglet levels caused by the eight-quark interaction thus 
lowers the value of $m_{f_0(600)}$ by about $200\,\mbox{MeV}$.   

There is no marked ef\mbox{}fect on the $K_0^*(800)$ state, which 
continues to lie above the $a_0(980)$ mass. The large $N_c$ result 
(\ref{K0*-a0}) protects the inequality $m_{K^*_0}>m_{a_0}$ for the
members of the quark-antiquark octet, which finally holds in the 
general case.      

To summarize, the ef\mbox{}fect of eight-quark interactions on the
mass spectrum, vacuum decay couplings, and mixing angles is relatively
small as long as general properties of the QCD vacuum (the values of 
the quark condensates and the topological susceptibility) are
correctly reproduced. 

%%%%%%%%%%%%%%%%%%%%%%%%%%%%%%%%%%%%%%%%%%%%%%%%%%%%%%%%%%%%%%%%%%%%%%%%%%%%%%
%
\section*{\centerline{\large\bf 6. Conclusions}}
%
%%%%%%%%%%%%%%%%%%%%%%%%%%%%%%%%%%%%%%%%%%%%%%%%%%%%%%%%%%%%%%%%%%%%%%%%%%%%%%
The role played by eight-quark interactions in the long wavelength
limit of QCD has been addressed in a systematic way. A full 
understanding of its impact on the vacuum and  properties of the 
low-lying spin zero mesonic spectra has been achieved. As an 
important by-product also the results associated with the well known 
four and six quark Lagrangians due to Nambu -- Jona-Lasinio and 't
Hooft, on the body of which the eight-quark terms are attached, are 
classified and presented according to stability criteria of the
effective potential (reviewed and illustrated in Section 2.2), the large 
$N_c$ counting scheme, approximate sum rules for meson masses, $U_A(1)$ 
and flavor $SU(3)$ breaking terms, and OZI-rule violation. This 
"dissection" allows not only for a complete understanding of the full 
result, (i.e. in the leading order stationary phase approximation to
the functional integral), for which we also give analytical
expressions, but also to compare with other relevant works in the
field. These discussions and respective formulae accompany every main 
derived step within the full result and serve as a "hitchhiker's
guide" to the mesons in the multi-quark "galaxy".\footnote{Borrowed
from "The ultimate hitchhiker's guide to the galaxy", by D. Adams.} 

An issue of much interest is the two-angle analysis of the $\eta
-\eta'$ mixing and its relation with the standard one-angle 
diagonalization. We clarify some existing confusion in the 
literature by deriving in full detail in Sections 3.2 and 3.3 the 
connection and equivalence of the two methods within the Lagrangian 
considered.

The masses and splitting of the complete result for $\eta -\eta'$
system are conveniently cast in the form of Eq. (\ref{Venext}), which 
separates in a transparent way the leading Gell-Mann -- Okubo 
contribution, the Witten -- Veneziano correction and the second-order 
$SU(3)$ flavor breaking corrections due to six and eight-quark terms. 
These latter ones have a positive sign for the parameter sets which 
yield good fits for the remaining pseudoscalar observables and
therefore add a small correction to the already large Witten -- 
Veneziano term, thus yielding a larger splitting than the empirically 
observed. 

Concerning the scalar sector we show that there exists a mass 
hierarchy  within the model considerations which is not conform with the 
present understanding of the empirical results. 

These drawbacks, being the result of an exhaustive and consistent 
study, clearly indicate that effects not considered, such as meson 
loops, higher orders in the heat kernel expansion, and confining 
forces, might be at work. 

We view therefore the main role of eight-quark forces considered as 
folows: (i) they are of vital importance for the stability of the 
ground state built from four and six-quark interactions. They restrict 
the choice of the couplings $G,\kappa,g_1,g_2$ to the rather narrow 
window of combinations given by Eq. (\ref{ineq1}). This is important, 
since combinations outside the allowed range can at instances even 
yield a better spectrum for the pseudoscalar mesons alone. It would be 
an erroneous result, attributing minor importance to the corrections
of the kind not considered. (ii) They help in understanding the
effects caused by the OZI-violating terms with coupling strength
$g_1$, which affect quite strongly the splitting of the $f_0^-,f_0^+$ 
scalars, mainly pushing down the lower state, due to the strong 
cancellations reported in Eq. (\ref{f0approx}). (iii) They may be also 
of importance in decays and scattering, not considered so far. (iv) 
They give a clear indication that an hierarchy of multi-quark 
interactions, with dominance of lower ones, is present. This 
corroborates with recent lattice calculations \cite{Bali} of gluon 
correlators, where the lower ones also dominate.

%%%%%%%%%%%%%%%%%%%%%%%%%%%%%%%%%%%%%%%%%%%%%%%%%%%%%%%%%%%%%%%%%%%%%%%%%%%%%%%
\section*{Acknowledgements}
%%%%%%%%%%%%%%%%%%%%%%%%%%%%%%%%%%%%%%%%%%%%%%%%%%%%%%%%%%%%%%%%%%%%%%%%%%%%%%%
This work has been supported by grants provided by Funda\c c\~ao para
a Ci\^encia e a Tecnologia, POCTI/FNU/50336/2003 and
POCI/FP/63412/2005. This research is part of the EU integrated 
infrastructure initiative Hadron Physics project under contract 
No.RII3-CT-2004-506078. A. A. Osipov also gratefully acknowledges 
the Funda\ca o Calouste Gulbenkian for f\mbox{}inancial support.

%%%%%%%%%%%%%%%%%%%%%%%%%%%%%%%%%%%%%%%%%%%%%%%%%%%%%%%%%%%%%%%%%%%%%%%%%%%%%%

\section*{\normalsize Appendix A. Response equation (\ref{res})}

%%%%%%%%%%%%%%%%%%%%%%%%%%%%%%%%%%%%%%%%%%%%%%%%%%%%%%%%%%%%%%%%%%%%%%%%%%%%%%
An algebraic system of nonlinear equations (\ref{res}) has, in
general, more than one admissible solution. Our specific problem
requires only the knowledge of all isolated real roots. In order to 
find them suppose that the system has at least one real solution,
i.e., the set $\{h_a\}$. Let us now manipulate with the equations at hand in
such a way that we can guess some values from this set. If our guess
is correct, the system reduces to a smaller one, which finally can be 
solved. It may happen however that the smaller system is incomplete and 
therefore has a continuum of solutions. Such cases have to be
excluded, because they are possible if and only if some of the components 
$\Delta_a$ are constrained (see below). It should be recalled that 
$\Delta_a$ is a set of independent variables (for different values of 
$a$) which will be fixed only later with the help of the gap
equations. Thus any restriction on $\Delta_a$ at this stage leads to
an internal contradiction and the corresponding solutions must be
rejected. There is another reason to exclude a continuum of solutions, 
namely the stationary phase method cannot be applied for such a case. 

After these general remarks, let's consider important details. Eq. 
(\ref{res}) yields the following set
\begin{equation}
\label{h1}
   \mu h_1 + \nu (h_4h_6+h_5h_7) =0,
\end{equation}
\begin{equation}
\label{h2}
   \mu h_2 + \nu (h_5h_6-h_4h_7) =0,
\end{equation}
\begin{equation}
\label{h3}
   \Delta_3 + \mu h_3 +\frac{\nu}{2}\left(h_4^2+h_5^2
            - h_6^2-h_7^2\right)=0, 
\end{equation}
\begin{equation}
\label{h4}
   \rho h_4 + \nu (h_1h_6-h_2h_7) =0,
\end{equation}
\begin{equation}
\label{h5}
   \rho h_5 + \nu (h_1h_7+h_2h_6) =0,
\end{equation}
\begin{equation}
\label{h6}
   \sigma h_6 + \nu (h_1h_4+h_2h_5) =0,
\end{equation}
\begin{equation}
\label{h7}
   \sigma h_7 + \nu (h_1h_5-h_2h_4) =0,
\end{equation}
\begin{equation}
\label{h8}
   \Delta_8 +\frac{1}{2}\left(\rho +\sigma\right)h_8
   +\frac{\nu}{\sqrt{3}}\left[h_1^2+h_2^2+h_3^2
   -\frac{1}{2}\left(h_4^2+h_5^5+h_6^2+h_7^2\right)\right]=0,
\end{equation}
\begin{eqnarray}
\label{h0}
  &&\Delta_0 +\xi h_0 +\frac{\kappa}{32}\sqrt{\frac{2}{3}}
    \left(3h_0^2-h_a^2\right)
    +\frac{g_2}{\sqrt{2}}h_8\left(h_1^2+h_2^2+h_3^2
    -\frac{h_8^2}{3}\right)  \nonumber \\
  &&+\frac{g_2}{2\sqrt{2}}\left[
    \left(h_4^2+h_5^2\right)
    \left(\sqrt{3}h_3-h_8\right)
    -\left(h_6^2+h_7^2\right)
    \left(\sqrt{3}h_3+h_8\right)\right] \nonumber \\
  &&+g_2\sqrt{\frac{3}{2}}\left[ h_1\left(h_4h_6+h_5h_7\right)
    + h_2\left(h_5h_6-h_4h_7\right)\right]=0,
\end{eqnarray}
where $\mu , \nu , \rho , \sigma$ and $\xi$ are defined as follows
\begin{equation}
\label{mu}
   \mu = G+\frac{\kappa}{8\sqrt{6}}\left(\sqrt{2}h_8-h_0\right)
       +\frac{1}{2}\left(g_1+g_2\right)h_a^2 + \frac{g_2}{2}h_0
       \left(h_0 + 2\sqrt{2}h_8\right),
\end{equation}
\begin{equation}
\label{nu}
   \nu = \frac{\kappa}{16} + g_2\sqrt{\frac{3}{2}}h_0, \qquad
   \xi = G +\frac{1}{2}\left(g_1+2g_2\right)h_a^2
         -\frac{2g_2}{3}h_0^2,
\end{equation}
\begin{equation}
\label{rho}
   \rho = G -\frac{\kappa}{16}\left(\sqrt{\frac{2}{3}}h_0
        - h_3 + \frac{h_8}{\sqrt{3}}\right)
        + \frac{1}{2}\left(g_1+g_2\right) h_a^2 
        + \frac{g_2}{2} h_0
       \left(h_0 + \sqrt{6}h_3 - \sqrt{2} h_8\right),
\end{equation}
\begin{equation}
\label{sigma}
   \sigma = G -\frac{\kappa}{16}\left(\sqrt{\frac{2}{3}}h_0
        + h_3 + \frac{h_8}{\sqrt{3}}\right)
        + \frac{1}{2}\left(g_1+g_2\right) h_a^2 
        + \frac{g_2}{2} h_0
       \left(h_0 - \sqrt{6}h_3 - \sqrt{2} h_8\right).
\end{equation}

Due to these definitions one has 
\begin{equation}
\label{mu-sigma}
   \mu -\sigma = \nu \left(\sqrt{3}h_8+h_3\right), \quad
   \mu -\rho   = \nu \left(\sqrt{3}h_8-h_3\right). 
\end{equation}

We start from the observation: if $\mu\neq 0$, one can multiply 
Eqs. (\ref{h4})-(\ref{h7}) by $\mu$ and, using 
Eqs. (\ref{h1})-(\ref{h2}), obtain 
\begin{eqnarray}
\label{4-7}
  &&h_4\left[\mu\rho -\nu^2\left(h_6^2+h_7^2\right)\right]=0,  \\
  &&h_5\left[\mu\rho -\nu^2\left(h_6^2+h_7^2\right)\right]=0,  \\
  &&h_6\left[\mu\sigma -\nu^2\left(h_4^2+h_5^2\right)\right]=0, \\
  &&h_7\left[\mu\sigma -\nu^2\left(h_4^2+h_5^2\right)\right]=0.
\end{eqnarray}
This leads to four possible alternatives:
\begin{eqnarray}
  &&1.\ \ \mu\neq 0, \quad h_4=h_5=h_6=h_7=0. \\
  &&2.\ \ \mu\neq 0, \quad h_4=h_5=0, \quad \sigma =0. \\
  &&3.\ \ \mu\neq 0, \quad h_6=h_7=0, \quad \rho =0.   \\
  \label{case4}
  &&4.\ \ \mu\neq 0, \quad \mu\rho =\nu^2\left(h_6^2+h_7^2\right), 
                      \ \ \mu\sigma =\nu^2\left(h_4^2+h_5^2\right). 
\end{eqnarray}

If $\mu =0$, then Eqs. (\ref{h1})-(\ref{h2}) now imply other four 
alternatives: 
\begin{eqnarray}
  &&5.\ \ \mu =0, \quad \nu =0. \\
  &&6.\ \ \mu =0, \quad \nu\neq 0,\quad  h_4=h_5=0. \\
  &&7.\ \ \mu =0, \quad \nu\neq 0,\quad  h_6=h_7=0. \\
  &&8.\ \ \mu =0, \quad \nu\neq 0,\quad  h_4=h_5=h_6=h_7=0. 
\end{eqnarray}

Cases 2 and 3, as well as cases 6 and 7, are correlated. The 
existence of such correlation becomes clear if one notes the
invariants of Eqs. (\ref{h1})-(\ref{h0}) under the substitutions
$h_4\leftrightarrow h_7$, $h_5\leftrightarrow h_6$, 
$h_3\leftrightarrow -h_3$, $\Delta_3\leftrightarrow -\Delta_3$.
In particular these substitutions change $\rho\leftrightarrow\sigma$,
leaving $\mu$ and $\nu$ without changes. 

{\it Case 1.} In this case Eqs. (\ref{h1})-(\ref{h2}) yield $h_1=h_2=0$
and one obtains a set of three equations to determine $h_0,h_3,h_8$.
To make further progress one should switch to the flavour basis 
$0,3,8\to u,d,s$, where one has
\begin{equation}
\label{SPA}
   \left\{ \begin{array}{l}
\vspace{0.2cm}   
   Gh_u + \Delta_u +\displaystyle\frac{\kappa}{16}\ h_dh_s
   +\displaystyle\frac{g_1}{4}\ h_u(h_u^2+h_d^2+h_s^2)
   +\displaystyle\frac{g_2}{2}\ h_u^3=0, \\
\vspace{0.2cm}   
   Gh_d + \Delta_d +\displaystyle\frac{\kappa}{16}\ h_uh_s
   +\displaystyle\frac{g_1}{4}\ h_d(h_u^2+h_d^2+h_s^2)
   +\displaystyle\frac{g_2}{2}\ h_d^3=0, \\
\vspace{0.2cm}   
   Gh_s + \Delta_s +\displaystyle\frac{\kappa}{16}\ h_uh_d
   +\displaystyle\frac{g_1}{4}\ h_s(h_u^2+h_d^2+h_s^2)
   +\displaystyle\frac{g_2}{2}\ h_s^3=0. 
   \end{array} \right.
\end{equation}
This system has been studied in our work \cite{Osipov:2005b}.

{\it Case 2.} 
Again Eqs. (\ref{h1})-(\ref{h2}) yield $h_1=h_2=0$. Then 
Eqs. (\ref{h3}) and (\ref{h8}) give $\Delta_3=\sqrt{3}\Delta_8$
(with the use of Eqs. (\ref{mu-sigma}) and $\sigma =0$). 
Since the variables $\Delta_a$ are supposed to be independent, one
concludes that this result leads to an apparent contradiction.
It may be also noted that the pair of quantities $(\Delta_3, \Delta_8)$
 define in general a plane in the nonet space, if neither quantity
vanishes. The found correlation means that only one axis is defined,
in which case the system is not complete and the solution is 
underdetermined. 

{\it Case 3.} 
As before Eqs. (\ref{h1})-(\ref{h2}) yield $h_1=h_2=0$. Then 
Eqs. (\ref{h3}) and (\ref{h8}) with the use of Eqs. (\ref{mu-sigma}) 
and $\rho =0$ give $\Delta_3=-\sqrt{3}\Delta_8$. One comes anew to 
the abovmentioned contradiction. 

{\it Case 4.} 
From Eqs. (\ref{case4}) one obtains $\mu (\rho -\sigma ) = \nu^2 (h_6^2
+h_7^2-h_4^2-h_5^2)$. Next, note that one can use Eqs. (\ref{mu-sigma})
to rewrite the result as follows 
$$
   \nu\left[2\mu h_3+\nu\left(h_4^2+h_5^2-h_6^2-h_7^2\right)\right]=0.
$$
Since $\nu\neq 0$ this yields $\Delta_3=0$. (The equality $\nu=0$
would lead here to $\rho =\sigma =0$, and finally due to 
Eqs. (\ref{mu-sigma}) to $\mu =0$. The obtained contradiction proves 
that $\nu\neq 0$.) Eqs. (\ref{h1})-(\ref{h2}), as well as equations
(\ref{case4}), imply that $\rho\sigma =\nu^2(h_1^2+h_2^2)$. Using 
this result and Eqs. (\ref{mu-sigma}), one can show that $\Delta_8=0$.
It is clear that in this case the solutions suffer from the same kind
of defects just mentioned above.    

{\it Case 5.} 
Since $\mu =\nu =0$, Eq. (\ref{mu-sigma}) gives $\rho =\sigma =0$. This
yields $\Delta_3=0$ and $\Delta_8=0$. This system is incomplete. 

{\it Case 6.}
Eqs. (\ref{h6})-(\ref{h7}) yield $\sigma =0$. Since $\mu =\sigma =0$,
Eq. (\ref{mu-sigma}) gives $h_3+\sqrt{3}h_8=0$. If $h_6$ and $h_7$ 
are nonzero, then from Eqs. (\ref{h4}) and (\ref{h5}), $h_1^2+h_2^2=0 
\to h_1=h_2=0$, since we only allow real solutions. As a result, one
obtains
$$
   \Delta_3-\sqrt{3}\Delta_8=\nu h_3\left(h_3+\sqrt{3}h_8\right)=0.
$$

{\it Case 7.} 
Eqs. (\ref{h4}) and (\ref{h5}) yield $\rho =0$. Next, due to $\sigma =0$,
Eqs. (\ref{h6}) and (\ref{h7}) reduce to $h_1h_4+h_2h_5=0$, 
$h_1h_5-h_2h_4=0$, it follows then $h_1=h_2=0$. Eqs. (\ref{mu-sigma}) 
give $h_3=\sqrt{3}h_8$, that finally leads to the correlation 
$\Delta_3=-\sqrt{3}\Delta_8$. 

{\it Case 8.} 
Eq. (\ref{h3}) reduces to $\Delta_3=0$. Eq. (\ref{h8}) has the form
$$
  \Delta_8 +\frac{\nu}{\sqrt{3}}
  \left(h_1^2+h_2^2+h_3^2-3h_8^2\right)=0.
$$
This case gives a new class of solutions. Nevertheless it can also be
thrown out, as long as $\Delta_3=0$, by the same arguments as just 
given.

One concludes that only the case 1 leads to the isolated real solutions
of Eqs. (\ref{res}) which must be taken into account. As has been shown
in \cite{Osipov:2005b} one can choose the parameters of the model in such
a way that only one real solution appears. 

We would like to note that our investigation here recalls in many aspects
the old results of Pais \cite{Pais:1968}, who studied the
response equations with octet driving forces. He dealt with the
octet space of $SU(3)$. Eqs. (\ref{res}) have some more complicated 
structure, but the general conclusions remain true. Even the formal 
covariance property of the system under the transformation
\begin{eqnarray}
  && h_0=q_0,\ h_1=-q_6,\ h_2=q_7,\ 
     h_3=\frac{1}{2}\left(q_3-\sqrt{3}q_8\right), \
     h_4=q_4,\ h_5=q_5, \nonumber \\
  && h_6=q_1,\  h_7=-q_2,\ 
     h_8=-\frac{1}{2}\left(q_8+\sqrt{3}q_3\right), 
\end{eqnarray}
in the nonet space is fulfilled. This transformation brings 
Eq. (\ref{res}) in the form
\begin{equation}
\label{resQ}
   b_a + Gq_a + \frac{3\kappa}{32} A_{abc} q_bq_c 
   +\frac{g_1}{2} q_a q_b^2 +\frac{g_2}{2} d_{abe}d_{cde} q_bq_cq_d=0,
\end{equation}
where
\begin{equation}
     b_3=\frac{1}{2}\left(\Delta_3-\sqrt{3}\Delta_8\right), \quad
     b_8=-\frac{1}{2}\left(\Delta_8+\sqrt{3}\Delta_3\right),
\end{equation}
and all other $b_a=0$.
 
%%%%%%%%%%%%%%%%%%%%%%%%%%%%%%%%%%%%%%%%%%%%%%%%%%%%%%%%%%%%%%%%%%%%%%%%%%%%%%

\section*{\normalsize Appendix B. Solving equations (\ref{hab1})
          and (\ref{hab2}) in the isotopic limit}

%%%%%%%%%%%%%%%%%%%%%%%%%%%%%%%%%%%%%%%%%%%%%%%%%%%%%%%%%%%%%%%%%%%%%%%%%%%%%%
As it has been shown in Appendix A, the system (\ref{res}) is actually
reduced to three coupled equations (\ref{SPA}) for three independent
variables $h_u, h_d, h_s$. These variables are defined as follows 
$h_a\lambda_a=\mbox{diag}(h_u,h_d,h_s)$, and easily related with 
$h_0, h_3$ and $h_8$, namely
\begin{eqnarray}
\label{fbases}
   && h_u=\frac{1}{\sqrt{3}}\left(\sqrt{2}h_0+\sqrt{3}h_3+h_8\right), 
      \nonumber \\
   && h_d=\frac{1}{\sqrt{3}}\left(\sqrt{2}h_0-\sqrt{3}h_3+h_8\right), 
      \nonumber \\
   && h_s=\sqrt{\frac{2}{3}}\left(h_0-\sqrt{2}h_8\right).
\end{eqnarray}

If we choose the current quark masses suitably, $m_u=m_d\neq m_s$, the 
f\mbox{}lavour symmetry of the action is broken down to the isospin 
group $SU(2)$. In this partial case one finds from Eqs. (\ref{res})
that $h_a=0$ for $a=1,2,3,4,5,6,7$, i.e., only two components $h_0$ and
$h_8$ are generally nonzero. Thus, in virtue of Eq. (\ref{fbases}) one
concludes that $h_u = h_d \neq h_s$. We shall suppose that $h_u$ and
$h_s$ are known\footnote{See \cite{Osipov:2005b} for details, where a 
more general case, $h_u\neq h_d\neq h_s$, has been considered. The 
case studied here is a straightforward consequence of that result.}. 

Let us solve Eqs. (\ref{hab1})-(\ref{hab2}) to find the couplings 
$h_{ab}^{(1)}$ and $h_{ab}^{(2)}$ for the case with isospin
symmetry. To represent the result we shall use dimensionless 
quantities 
\begin{equation}
   \omega_i  =\frac{\kappa h_i}{16G}\, ,\qquad
   \rho_{ij} =\frac{g_1 h_ih_j}{4G}\, , \qquad
   \tau_{ij} =\frac{g_2 h_ih_j}{2G}\, ,
\label{rhoomtau}
\end{equation}
induced by six- and eight-quark interactions with strength couplings 
$\kappa, g_1$ and $g_2$ correspondingly. We use also that 
$\rho =2\rho_{uu}+\rho_{ss}$. 

The result for $h_{ab}^{(1)}$ is  
\begin{eqnarray}
   h_{ab}^{(1)}= \left\{
   \begin{array}{lcl}
   \displaystyle \frac{-\delta_{ab}}{G(1-\omega_s 
               +\rho +3\tau_{uu})} 
   &\quad & (a,b =1,2,3),\\
   \displaystyle \frac{-\delta_{ab}}{G(1-\omega_u 
               +\rho +\tau_{uu} +\tau_{ss}+\tau_{us})} 
   &\quad & (a,b =4,5,6,7),\\
   \displaystyle \frac{-N^{(1)}_{ab}}{G \det N^{(1)}} 
   &\quad & (a,b =0,8). \\
   \end{array} \right.
\end{eqnarray}
where the $2\times 2$ symmetric matrix $N^{(1)}$ has elements
\begin{eqnarray}
   && N^{(1)}_{00} = 1+\frac{1}{3}\left(\omega_s -4\omega_u 
      +10\rho_{uu}+7\rho_{ss} -8\rho_{us}\right) 
      +\tau_{uu} +2\tau_{ss}, 
      \nonumber\\
   && N^{(1)}_{08} = N^{(1)}_{80} =\frac{\sqrt{2}}{3}\left[
      \omega_u-\omega_s +2(\rho_{ss} -2\rho_{uu}+\rho_{us})
      +3(\tau_{ss}-\tau_{uu})\right], 
      \nonumber \\
   && N^{(1)}_{88} = 1+\frac{1}{3}\left(4\omega_u +2\omega_s 
      +14\rho_{uu}+5\rho_{ss} +8\rho_{us}\right) 
      +2\tau_{uu} +\tau_{ss},
\end{eqnarray}
and its determinant is equal to
\begin{eqnarray}
   \det N^{(1)}&\!\! =\!\! & 1 + \omega_s - 2\omega_u^2 + 4\rho 
   + 3\left[(1+\rho )(\tau_{uu}+\tau_{ss}) +\rho^2 
   \right. \nonumber \\
   &\!\! +\!\!&\left. \omega_s (\rho_{ss}-2\rho_{uu}+\tau_{ss})
   +3\tau_{us}^2+ 6\tau_{us}\rho_{us}\right].
\end{eqnarray}

For $h_{ab}^{(2)}$ one obtains  
\begin{eqnarray}
   h_{ab}^{(2)}= \left\{
   \begin{array}{lcl}
   \displaystyle \frac{-\delta_{ab}}{G(1+\omega_s 
               +\rho +\tau_{uu})} 
   &\quad & (a,b =1,2,3),\\
   \displaystyle \frac{-\delta_{ab}}{G(1+\omega_u 
               +\rho +\tau_{uu} +\tau_{ss}-\tau_{us})} 
   &\quad & (a,b =4,5,6,7),\\
   \displaystyle \frac{-N^{(2)}_{ab}}{G \det N^{(2)}} 
   &\quad & (a,b =0,8). \\
   \end{array} \right.
\end{eqnarray}
where the $2\times 2$ symmetric matrix $N^{(2)}$ has elements
\begin{eqnarray}
   && N^{(2)}_{00} = 1+\frac{1}{3}\left(4\omega_u -\omega_s\right)
      +\rho + \frac{1}{3}\left(\tau_{uu} +2\tau_{ss} \right), 
      \nonumber\\
   && N^{(2)}_{08} = N^{(1)}_{80} =\frac{\sqrt{2}}{3}\left(
      \omega_s-\omega_u 
      +\tau_{ss}-\tau_{uu}\right), 
      \nonumber \\
   && N^{(2)}_{88} = 1-\frac{2}{3}\left(2\omega_u +\omega_s\right) 
      +\rho +\frac{1}{3}\left(2\tau_{uu} +\tau_{ss}\right),
\end{eqnarray}
and 
\begin{equation}
   \det N^{(2)} = (1+\rho)(1+\rho +\tau_{uu} + \tau_{ss}-\omega_s)
   -2\omega^2_u +\tau_{ss}(\tau_{uu}-\omega_s).
\end{equation}

%%%%%%%%%%%%%%%%%%%%%%%%%%%%%%%%%%%%%%%%%%%%%%%%%%%%%%%%%%%%%%%%%%%%%%%%%%%%%

\end{document}